\documentclass[letterpaper]{article} 
\usepackage{aaai25}  
\usepackage{times}  
\usepackage{helvet}  
\usepackage{courier}  
\usepackage[hyphens]{url}  
\usepackage{graphicx} 
\usepackage{natbib}  
\usepackage{caption} 
\usepackage{algorithm}
\usepackage{algorithmic}
\usepackage{amsfonts,amssymb,amsmath}
\usepackage{subcaption}
\usepackage{mathrsfs}
\usepackage{newfloat}
\usepackage{listings}
\DeclareCaptionStyle{ruled}{labelfont=normalfont,labelsep=colon,strut=off} 
\floatstyle{ruled}
\newfloat{listing}{tb}{lst}{}
\floatname{listing}{Listing}
\title{Contrastive Representation for Interactive Recommendation}

\title{Contrastive Representation for Interactive Recommendation}
\author {
    Jingyu Li\textsuperscript{\rm 1},
    Zhiyong Feng\textsuperscript{\rm 1},
    Dongxiao He\textsuperscript{\rm 1},
    Hongqi Chen\textsuperscript{\rm 1},
    Qinghang Gao\textsuperscript{\rm 1},
    Guoli Wu\textsuperscript{\rm 1}
}
\affiliations {
    \textsuperscript{\rm 1}College of Intelligence and Computing, Tianjin University\\
    \{lijingyu\_working, zyfeng, hedongxiao, hongqi, gaoqh, wuguoli\_it999\}@tju.edu.cn
}

\usepackage{bibentry}

\begin{document}

\maketitle

\begin{abstract}
Interactive Recommendation (IR) has gained significant attention recently for its capability to quickly capture dynamic interest and optimize both short and long term objectives. IR agents are typically implemented through Deep Reinforcement Learning (DRL), because DRL is inherently compatible with the dynamic nature of IR. However, DRL is currently not perfect for IR. Due to the large action space and sample inefficiency problem, training DRL recommender agents is challenging. The key point is that useful features cannot be extracted as high-quality representations for the recommender agent to optimize its policy. To tackle this problem, we propose Contrastive Representation for Interactive Recommendation (CRIR). CRIR efficiently extracts latent, high-level preference ranking features from explicit interaction, and leverages the features to enhance users' representation. Specifically, the CRIR provides representation through one representation network, and refines it through our proposed Preference Ranking Contrastive Learning (PRCL). The key insight of PRCL is that it can perform contrastive learning without relying on computations involving high-level representations or large potential action sets. Furthermore, we also propose a data exploiting mechanism and an agent training mechanism to better adapt CRIR to the DRL backbone. Extensive experiments have been carried out to show our method's superior improvement on the sample efficiency while training an DRL-based IR agent.
\end{abstract}

%

\section{Introduction}
Interactive Recommendation (IR) is recently popular and becoming accepted as an reasonable recommender workflow. Traditionally, the recommendation problem was considered to be a classification or prediction task (such us collaborative filtering and content-based filtering methods). However, this may not match the real recommendation scenario. It is now widely agreed that formulating it as a sequential decision problem can better reflect the user-system interaction \cite{RLSurvey}. Therefore, IR can be formulated as a Markov decision process and be solved by Reinforcement Learning (RL) or Deep Reinforcement Learning (DRL). DRL-based IR can naturally capture users’ unique dynamic interests and balance between short and long term targets, similar to the well-known Reinforcement Learning based on Human Feedback (RLHF) mechanism in ChatGPT \cite{openai2024gpt4}. There has been a variety of commercial services on interactive recommendation systems based on DRL \cite{youtubeRLRec, VisualDialog, huaweiKGRL, kuaishouRLUR}. 

However, sample inefficiency is a significant issue that hinders the further development of IR \cite{ijcai2018p820}. Sample efficiency refers to the training performance which can be achieved limited to a certain number of training samples. It measures the training difficulty of an RL agent. For IR tasks, the DRL models usually turn out to be even more sample-inefficient than other typical DRL tasks (e.g., robotic control, game agents, etc.) \cite{Chen2021ASO}. Because conducting DRL from high dimensional observations is empirically observed to be sample-inefficient \cite{lake2017building, kaiser2024modelbased}. Unfortunately, IR usually has to encode users' profiles into high dimensional observations to convey abundant semantic information. This makes the IR agents can be hardly trained to the ideal effect within limited online interaction. So it cannot quickly attract users' interest, leading to the failure of maintaining a certain number of active users \cite{CIRS}. This is a fatal problem for the online recommendation business.

Some approaches have been proposed to address the sample efficiency problem in IR. Commonly, they can be classified into three streams of methods based on different intention (some methods may belong to more than one category): \textbf{(i)} Improve functional components in DRL; \textbf{(ii)} Increase significative reward signals. \textbf{(iii)} Enhancing the state representation method. The first class enhance the policy for making action \cite{NICF} or the way of exploiting samples \cite{LSER}. The second class usually trains off-line user simulator to simulate users' behaviours and give reward feedback towards recommendation \cite{Virtual-Taobao,ie2019recsim,rohde2018recogym,zhao2023kuaisim}. The last class aims at enhancing the representation methods for extracting users' profile \cite{DRR,CRRTrans}. Works of the last class are usually based on the consensus \cite{CURL}: \textit{If an agent can acquire high quality semantic information from high dimensional observations, DRL-based recommendation methods built on top of those representations should be significantly more sample-efficient}. In this paper we name it \textbf{DRL Representation Consensus}.

Our work falls into the last class of work, which refines the state representation. But rather than process state information feed-forwardly (such us pooling embeddings or applying a neural network), we consider to use an auxiliary task paralleled with the main DRL task to learn semantic information for representations. Our motivation comes from the self-supervised contrastive learning in traditional Deep-Learning recommendation paradigm. However, there are three obvious problems:
\textbf{(i)} In traditional recommendation paradigm, sufficient contrastive samples are derived from static datasets. But in IR scenarios, interaction history cannot provide such enough samples. 
\textbf{(ii)} In traditional recommendation paradigm, contrastive learning is usually used to constrain users' high-level sequence or graph representations. But directly applying it in IR will cost greatly for the large action space.
\textbf{(iii)} IR models conducts online recommendation and offline training simultaneously, so contrastive learning must be conducted along with online recommendation. Whether a stable IR agent will be successfully trained in this way has not been very clear.

To tackle these problems, we propose Contrastive Representation for Interactive Recommendation (CRIR) method. The CRIR is implemented through one state representation network and our proposed Preference Ranking Contrastive Learning (PRCL). The PRCL tackles the problem (i) by fully taking advantage of users' different preference measurements towards different interacted items at every moments. The state representation network addresses problem (ii) by generating interest weights to select behavior representations which approximate the high-level user representation. This approach along with PRCL could avoid computation around whole potential action set mentioned in problem (ii). Through ranking those interests weights, a Positional Weighted InfoNCE Loss in PRCL is applied to maximize the agreements between user's preferable interests at a specific moment. Different from prior contrastive methods in DRL(\cite{CURL,DBC}), we apply an data exploiting and agent training mechanism to solve problem (iii). In these two mechanisms, PRCL is conducted separately with main DRL task, but can achieve better effect. Extensive experiments conducted on Virtual-Taobao simulation environment and a simulator based on ml-1m dataset further verify the effectiveness of the whole proposed CRIR.

\section{Related Works}

\subsection{Interactive Recommendation}
Interactive recommendation is an online task in which agents generate recommended items and optimizes itself in the process of interacting with users. It usually models the recommendation problem as a Markov decision process and solved by RL or DRL \cite{RLSurvey, Chen2021ASO}. DRL is trained by a reward feedback evaluating its action towards the current state. But Traditional recommendation datasets are sparse and cannot give an explicit rating towards every action. So some researchers develop reward models which tracks and simulates users' behaviors from datasets or online services \cite{Virtual-Taobao,ie2019recsim,rohde2018recogym, zhao2023kuaisim}. Some researchers collect some dense datasets to ease further research \cite{kuairec, kuairand}.

IR has been studied from various standpoints. SlateQ \cite{SlateQ} was proposed to decompose slate Q-value to estimate a long-term value for individual items, stating a way to recommend a page-view of items through one interaction. PGCR \cite{PGCR} utilized both policy gradients, time-dependent greed and actor-dropout to balance exploration and exploitation. TPGR \cite{TPGR, TPGR+} designed a tree-structured policy gradient method to handle the large discrete action space hierarchically. Cai et al. \cite{Stochastic} designed two stochastic reward stabilization frameworks to replace the direct stochastic feedback with that learned by a supervised model so that to stabilize training process. In addition to general interactive recommendation, many scholars have paid attention to the practicability of IR systems. CIRS \cite{CIRS} designed a causal inference based model to burst Filter Bubbles in IR. DORL \cite{Matthew} made detailed analysis on Matthew Effect in IR and contribute to penalizes unbalanced exposure distribution. Dubbed RLUR \cite{kuaishouRLUR} focused on the user retention issue on short video IR.

\subsection{Contrastive Learning in Recommender System}
Contrastive Learning (CL) and Self-Supervised Learning (SSL) have brought much attentions by different research communities including CV \cite{simclr, moco} and NLP \cite{simcse, NLPCon}. Some works concentrated on applying CL or SSL in DRL \cite{CURL, DBC} but most of which were centralized on enhancing vision encoders for RL algorithms. As far as we concerned, few works have ever tried CL for IR paradigm. We make discussions mainly on contrastive self-supervised learning in recommender system.

Applying CL in sequential recommendation models raised much attentions in recent years \cite{SRSurvey}. Xin et al. \cite{Xin2020SelfSupervisedRL} used dataset labels to compute cross-entropy loss as reward to train a RL model, then used the RL model to enhance existing self-supervised sequential recommendation models in deep learning paradigm. GESU \cite{ICWS2022} concentrated on incorporating social information to sequential recommendation models. ICL  \cite{ICL} learns users’ intent distributions via clustering, and then leverages the learnt intents into the user representation via their proposed contrastive approach. Graph contrastive learning also performs well on graph based recommendation tasks \cite{GCLSurvey}. SGL \cite{SGL} adopted a multi-task framework with contrastive SSL to improve the GCN-based collaborative filtering methods \cite{LightGCN, NGCF}. NCL (Neighborhood-enriched Contrastive Learning) \cite{NCL} explicitly incorporates the potential semantic neighbors into contrastive pairs to enrich semantic information in graph. LightGCL \cite{LightGCL} can alleviate the problem caused by inaccurate self-supervised contrastive signals by injecting global collaboration.

\subsection{Sample Efficiency in IR}\label{related work}
As mentioned in the introduction, sample inefficiency is a tricky problem for DRL and IR \cite{Chen2021ASO} that still remains to be well treated. Many classical DRL methods also have many strategies to deal with sample inefficiency, e.g. PPO \cite{PPO}, SAC \cite{SAC1, SAC2}, CRR \cite{CRR}, DDPG \cite{DDPG}. However, those naive DRL methods are not enough to treat with IR scenarios. Many works have broadened new horizons to make interactive recommendation more reliable. DRR \cite{DRR} proposed some basic state representation method and a generative recommendation paradigm utilizing DDPG. NICF \cite{NICF} designed an exploration policy with multi-channel transformer to capture users' shifting interest in cold-start settings. KGRL \cite{huaweiKGRL} utilized knowledge graph to enhanced semantic information in reinforcement learning. Xi et al. \cite{CRRTrans} used transformer as state representation network and CRR as backbone RL framework along with pre-trained embeddings to make recommendation. LSER \cite{LSER} applied Locality-Sensitive Hashing algorithm in experience replay procedure to sample most valued training batches. DACIR \cite{sigir/WuX0ZZL22} aligned embeddings from different domain into a shared latent space to fertilize embedding information for cross-domain interactive tasks.

Although IR has gained significant attention recently, research on its sample efficiency remains neither sufficient nor systematic. Some recent works are noteworthy, but they address different problems or are applied in very different contexts (such as TPGR, DORL, KGRL, LSER, DACIR, etc.). Consequently, our options for baseline are limited. So we choose SAC, CRR, PPO, DRR, and NICF as baselines.

\section{Contrastive Representation}

\subsection{Framework Preliminaries}

\begin{figure*}[ht]
    \centering
    \includegraphics[width=\linewidth]{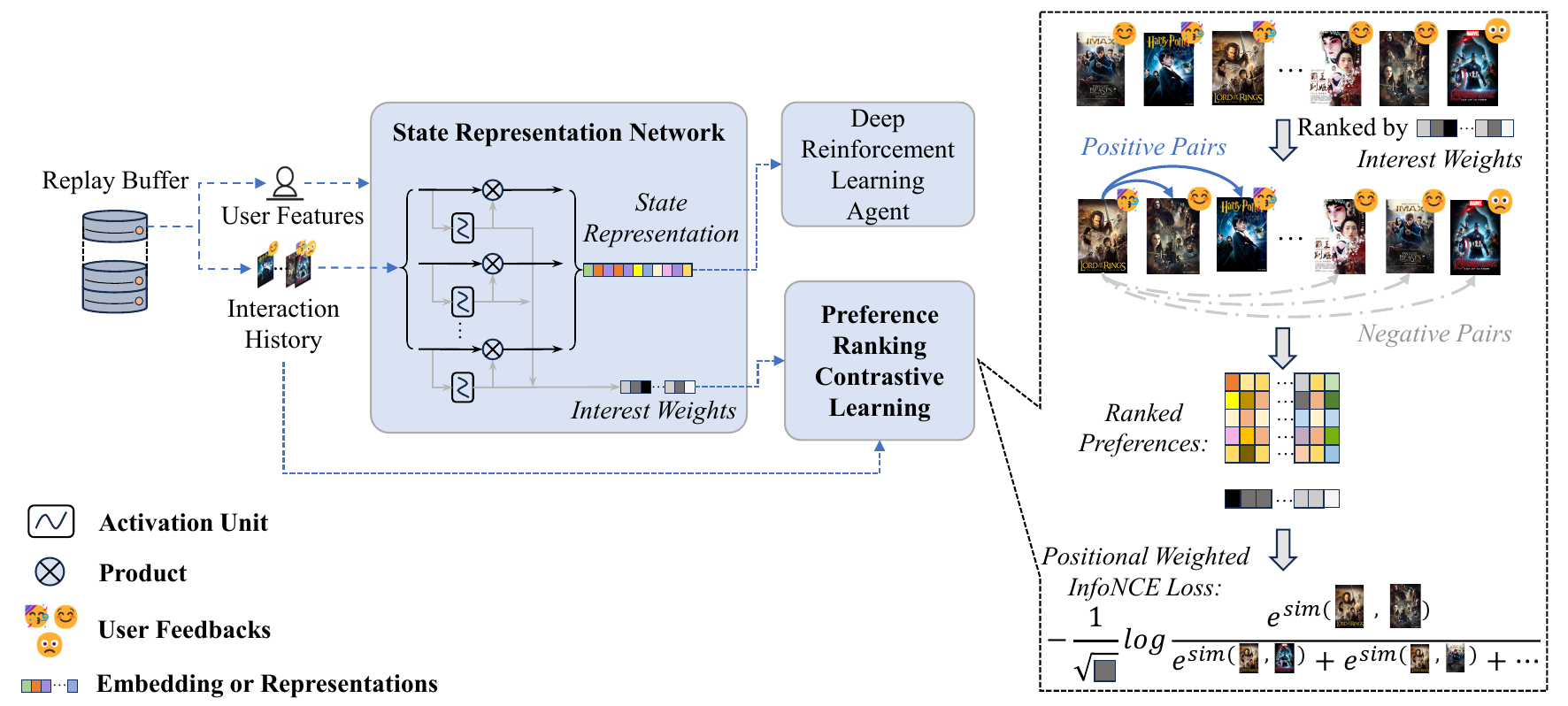}
    \caption{Overview of Contrastive Representation for Interactive Recommendation. }\label{framework}
\end{figure*}

CRIR uses the auxiliary, paralleled task PRCL to get better representations for main DRL task. In this paper we name this training mechanism as \textbf{Auxiliary Mechanism.} As shown in Figure~\ref{framework}, the proposed Contrastive Representation is composed of a State Representation Network and the PRCL method. They cooperate to acquire high level representations through the connection of \textit{Interest Weight}. The \textit{Interest Weight} is utilized to formulate \textit{State Representation} for RL, and also indicate the importance of the interacted items in PRCL at every specific moment. The replay buffer is a general components in off-policy DRL \cite{DDPG}. Here it stores historical interaction transitions. It will sample a batch of transitions while training the agent and conducting PRCL. Each transition contains one users' interaction history and other profiles at one past moment. In our implementation, we use DDPG  \cite{DDPG} along with Priority Experience Replay mechanism (PER) \cite{PER} as our DRL backbone for its effectiveness and stability.

\subsection{State Representation Network} \label{4.2}
Some works have already employ attention mechanism or transformer \cite{transformer} to model state representation in IR \cite{DRR, CIRS, CRRTrans}. However, 
what we need is the explicit degree of emphasis to different behaviors of the user at a specific timestamp. We discover that the weighted sum attention mechanism in Deep Interest Network \cite{zhou2018deep} naturally fits this paradigm. Its effectiveness is also validated by various online recommendation services. So it is determined as part of the state representation network to model the state information and generate preference scores of interacted items.

\begin{figure}[ht]
    \centering
    \includegraphics[width=\linewidth]{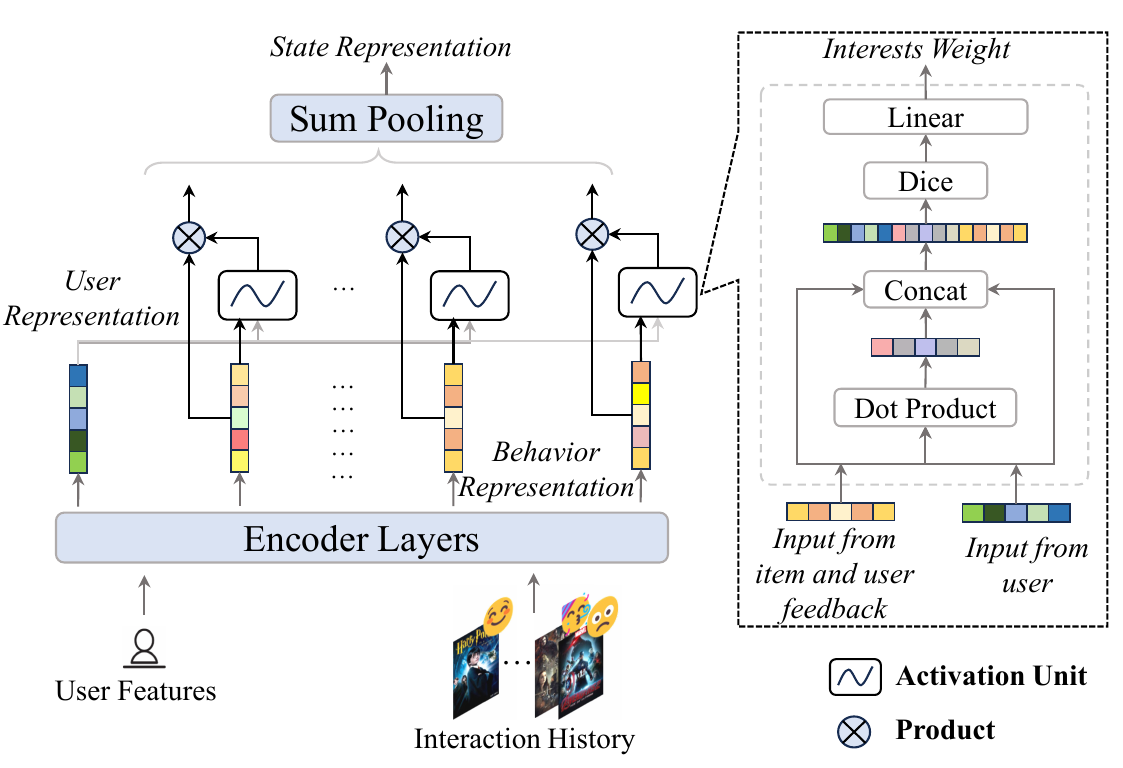}
    \caption{Weighted sum part of the state representation network ($\sum_{\tau=1}^{t}\Lambda(u_t, h_{\tau}) \cdot h_{\tau}$ in equation Eq~(\ref{SR})).}\label{state}
\end{figure}

As shown in Figure~\ref{state}, features and behavior histories of the current user are fed into their respective embedding layers. Behavior history contains not only item features but also feedback given by the user at each moment. Then weights for each single behavior will be computed through each activation unit, whose specific structure is shown in Figure~\ref{state}. The settings for activation unit and Dice activation function follow Deep Interest Network \cite{zhou2018deep}. 

Average information should be preserved to retain basic state information and stabilize convergence. This idea is proved to be effective in DRR \cite{DRR}. So we employ the average pooling in parallel with the weighted sum attention module. The final state representation of user $u$ at timestamp $t$ is formulated as:\begin{equation}\label{SR}
    s_{u,t}=(\frac{1}{t}\sum_{\tau=1}^{t}u_{t}\otimes h_{\tau})
    \oplus (\sum_{\tau=1}^{t}\Lambda(u_t, h_{\tau})\cdot h_{\tau}),
\end{equation}
where $\Lambda(\cdot, \cdot)\in \mathbb{R}$ is the activation unit, with representations of user $u_t\in \mathbb{R}^{D_R}$ and behaviors $h_{\tau}\in \mathbb{R}^{D_R}$ as input, $D_R$ is the representation dimension, $\otimes$ and $\oplus$ stands for the outer product and concatenation separately.

\subsection{Preference Ranking Contrastive Learning}
This section will specifically states our proposed PRCL. We will first give a brief introduction to the problem definition and then specifically introduce the procedure of PRCL, including Data Augmentation and Positional Weighted InfoNCE Loss. Figure~\ref{framework} could vividly show the process.

\subsubsection{IR Objective}
The optimization objective of the whole IR process could be formulated as maximizing $J_{\omega,\theta}$:
\begin{equation}
     J_{\omega,\theta}={\sum_{u=1}^{|\mathcal{U}|}\sum_{t=1}^{T_u}\mathbb{E}_{a_{t;\theta}}[r(s_{u,t;\omega},a_{u,t;\theta})]},
\end{equation}
where $\omega,\theta$ are the parameter sets for state representation and DRL components, separately.  $\mathcal{U}$ is the user set, $T_u$ is the interaction length for user $u$, $s_{t;\omega}\in \mathbb{R}^{D_S}$ is the representation for user state $s$ at timestamp $t$ with parameter set $\omega$.  $r(s_t,a_t)$ is the reward returned from environment while taking action $a_t$ at state $s_t$. The action $a_{t;\theta}\in \mathbb{R}^D$ is actually the representation of the chosen item to be recommended. This maximization goal could be transformed to the goal particularly for PRCL task around user $u$:
\begin{equation}\label{PRCLGoal}
J_{u;\omega}=\sum_{t=1}^{T_u}\sum_{k}^{|\mathcal{A}_{s_t}|}\ln{P(s_{t;\omega},a_{t,k})},
\end{equation}
where $P(s_{t;\omega},a_{t,k})$ is the joint probability for the agent taking action $a_{t,k}$ at state $s_{t}$, $\mathcal{A}_{s_t}$ is the potential action set for state $s_{t;\omega}$. Our PRCL is concentrated on optimizing Eq~(\ref{PRCLGoal}).

\subsubsection{Data Augmentation} \label{dataaug}

\textbf{}

\textbf{(i) Sampling}: The replay buffer samples history interactions to train the DRL agent. We should also sample data for PRCL. Considering that PRCL is conducted at different stage with the main DRL task, we design an data sampling mechanism, which can achieve both two optimization goals simultaneously. In our implementation, we use two batches of data to conduct contrastive learning. One is sampled in totally random from the replay buffer, and another is the data for training the DRL networks in the next round, using the PER sampling strategy. This mechanism ensures that every transition undergoing reinforcement learning also experiences contrastive learning at least once. In this paper we name this data exploiting mechanism as \textbf{Mixed Mechanism}. Therefore, the goal of DRL and PRCL could be achieved together although they are conducted separately. Our experiments will study the Mixed Mechanism specifically.

\textbf{(ii) Weighting}: As shown in Figure~\ref{state}, the state representation network can either model the state information or generate interest weights for different behaviors. One single behavior with larger weight value means that the current user is predicted to pay more attention to the item in this behavior. These weights plays critical roles at the following \textbf{(iii) Ranking} step and Positional Weight InfoNCE Loss.

\textbf{(iii) Ranking}: \label{4.3.3}Every interaction in the sampled batch is assigned with an interest weight as mentioned in \textbf{(ii) Weighting}. Suppose the length for an interaction history is $n$ with max sequence length $M$ ($n \le M$). The ranked list of the behavior representation sequence is formed as $[h_1, h_t, \cdots, h_n]$ with higher interest weight ranking ahead.

Then positive and negative pairs for contrastive learning should be generated. In every interaction, the attention scores ranking the second to the the $\lfloor n/2\rfloor$-th will be treated as candidate positive items. Randomly choose $k \in \{2,\cdots ,\lfloor n/2\rfloor\}$, then $(h_1, h_k)$ is treated as the positive pair. Every tuple like $(h_1, h_t)$ where $t \in  \{\lfloor n/2\rfloor+1,\cdots, n\}$ is treated as negative pairs for this interaction. Finally, we get one positive pair and $\lceil n/2\rceil$ negative pairs for each transition in the training batch to conduct PRCL.

\subsubsection{Positional Weight InfoNCE Loss}
Since it is reasonable for the agent to make action according to the current state, it's reasonable for this distribution of action $a_{t,k}$ in Eq~(\ref{PRCLGoal}) to be written as Gaussian-distribution-like loss around state $s_t$:
\begin{equation}
     \mathcal{L}_u=-\sum_{t=1}^{T_u}\sum_{k}^{|\mathcal{A}_{s_t}|}\log \frac{\exp(a_{t,k}^{\rm T}W s_{t})}{\sum_{j=1}^{|\mathcal{A}_{s_t}|}\exp(a_{t,j}^{\rm T}W s_{t})},
\end{equation}
where $W\in \mathbb{R}^{D\times D_S}$. In order to simplify the computational complexity, we utilize representative behavioral representation to approximate the state $s_t$ and the sum-up operation around potential action set $\mathcal{A}_{s_t}$. The optimization goal for user $u$ at timestamp $t$ could be formulated as:
\begin{equation}
         \mathcal{L}_u(t)=-\log \frac{\exp(h^{\rm T}_{k}\cdot h^{*})}{\sum_{n=1}^{|\mathcal{N}_{s_t}|}\exp(h_n^{\rm T} \cdot h^{*})},
\end{equation}
where $h^{*} =h_i,\ i=\mathop{\mathrm{argmax}}\limits_{i\leq t}\ w_i,\ t \leq T_u$ is the optimal behavioral representation, $k$ is the chosen behavior index from positive set mentioned in \textbf{(iii) Ranking}, $\mathcal{N}_{s_t}$ is the negative behavior set at current state $s_t$, also mentioned in \textbf{(iii) Ranking}. $\mathcal{N}_{s_t}$ could be seen as negative sampling, utilized to alternate computation on the whole potential action set.

Considering that different $h_k$ should have different similarity value with $h^{*}$, we decide to use a coefficient to model this discrimination of different contrastive pairs. Obviously the ranking position can measure the importance of the contrastive pair in training the representation. \textbf{So we use $\mathbf{1/\sqrt{R_u(h_k)}}$ to smooth the discrimination}, where $R_u(h_k)$ is the ranking position for item $h_k$ of user $u$ mentioned at \textbf{(iii) Ranking}. The proposed Positional Weighted InfoNCE Loss is formulated as:
\begin{equation}\label{rcl}
\begin{aligned}
        \mathcal{L}_u(t)=-\frac{1}{\sqrt{R_u(h_k)}}\log\frac{\exp(h^{\rm T}_k\cdot h^{*})}{\sum_{n}^{{\mathcal{N}}_{s_t}}\exp(h_n^{\rm T} \cdot h^{*})}.
\end{aligned}
\end{equation}

\section{Experiments}
In experiment section we want to investigate the following research questions.
\begin{itemize}
\item \textbf{(RQ1)} How does CRIR perform with other IR methods aiming at improving sample efficiency?
\item \textbf{(RQ2)} What contributions dose each PRCL components make in the whole system?
\item \textbf{(RQ3)} Does the sampling and training mechanism contribute greatly to the training performance?
\end{itemize}

\subsection{Experimental Setup}
\label{5.2}
\begin{figure*}
\centering
	\subcaptionbox{Episode reward for Virtual-Taobao}{\includegraphics[width = 0.33\textwidth]{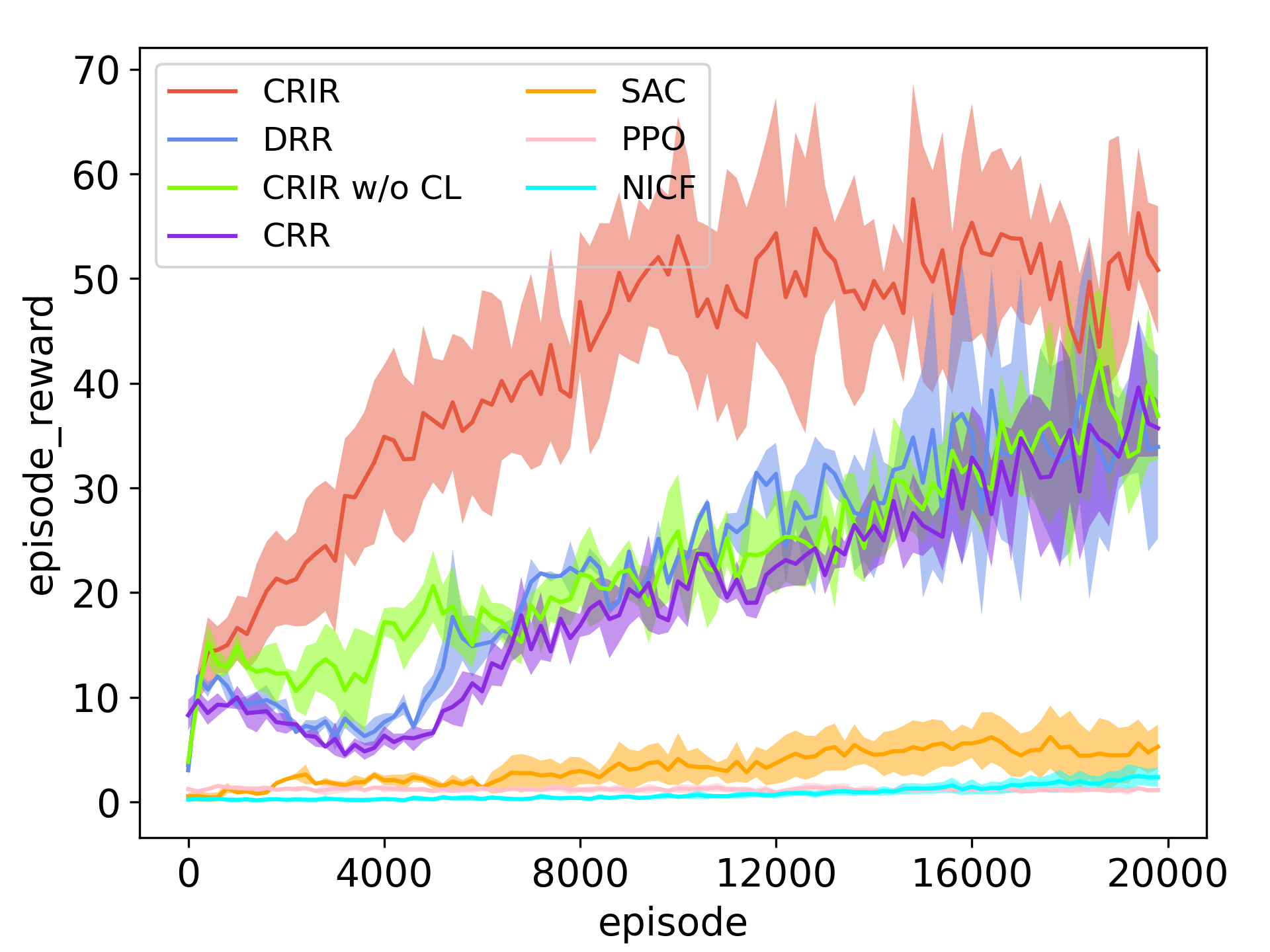}}
	\hfill
	\subcaptionbox{CTR for Virtual-Taobao}{\includegraphics[width = 0.33\textwidth]{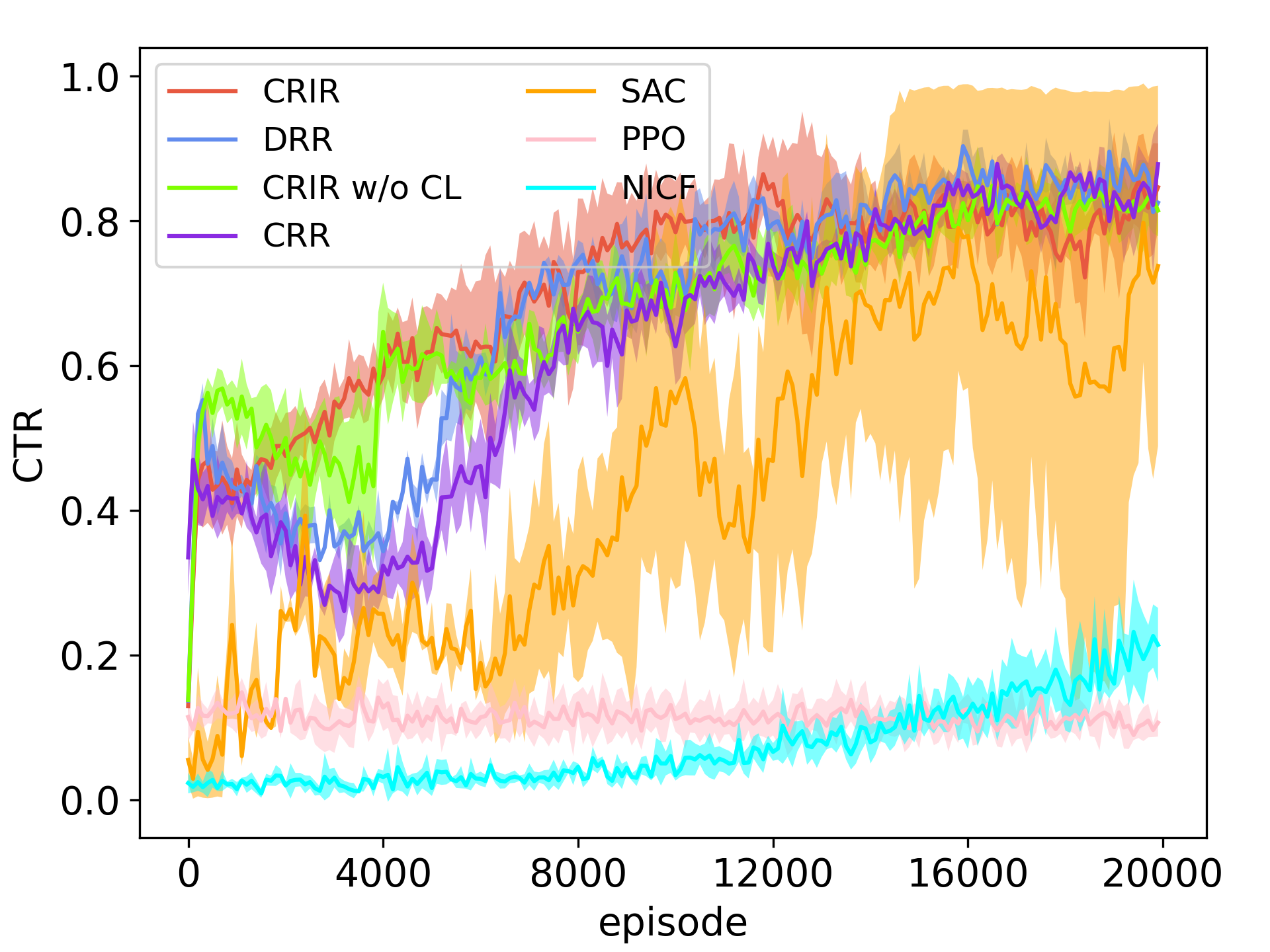}}
	\hfill
	\subcaptionbox{Episode reward for ML-1M}{\includegraphics[width = 0.33\textwidth]{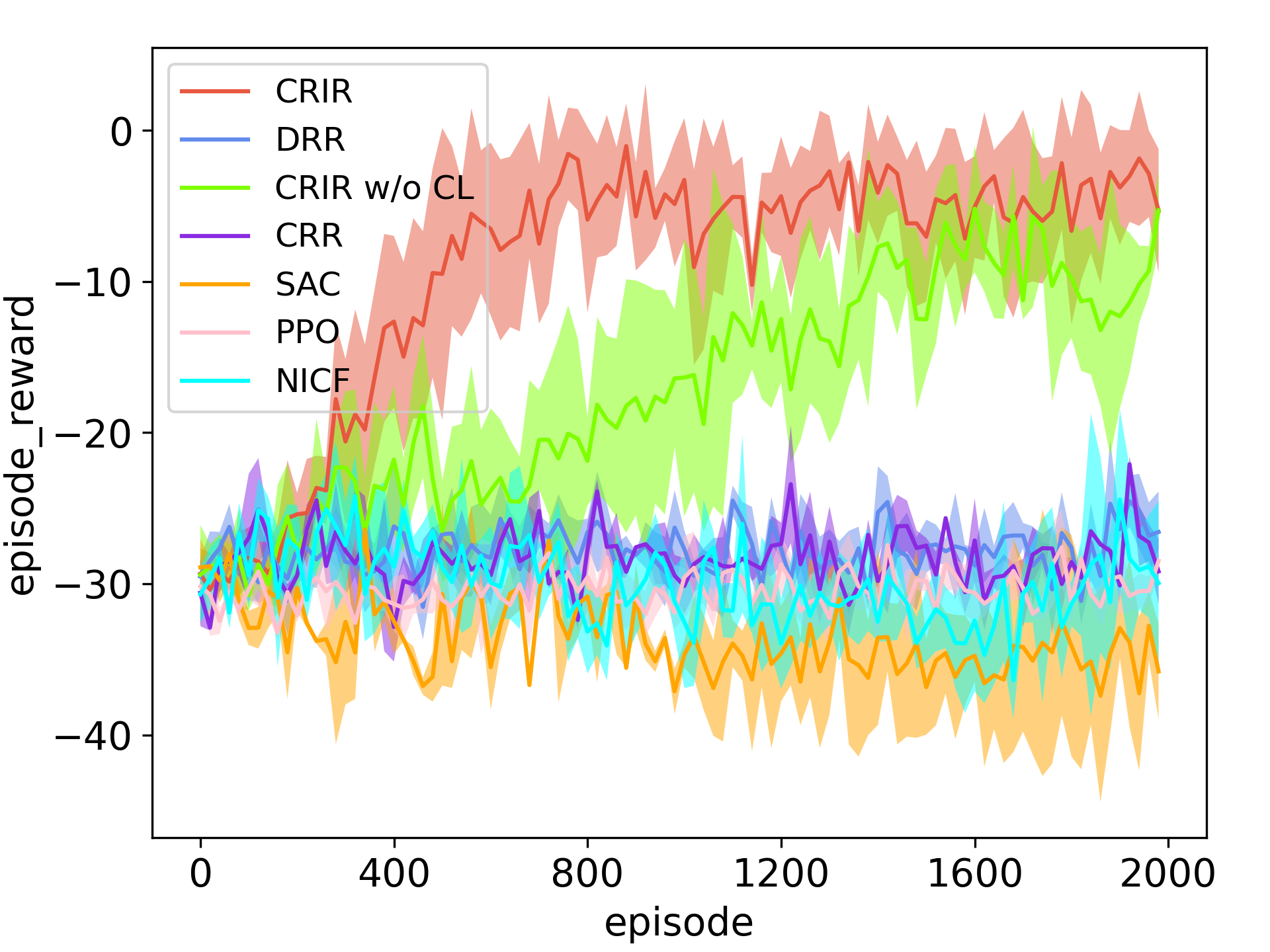}} 

        \subcaptionbox{CTR for ML-1M}{\includegraphics[width = 0.33\textwidth]{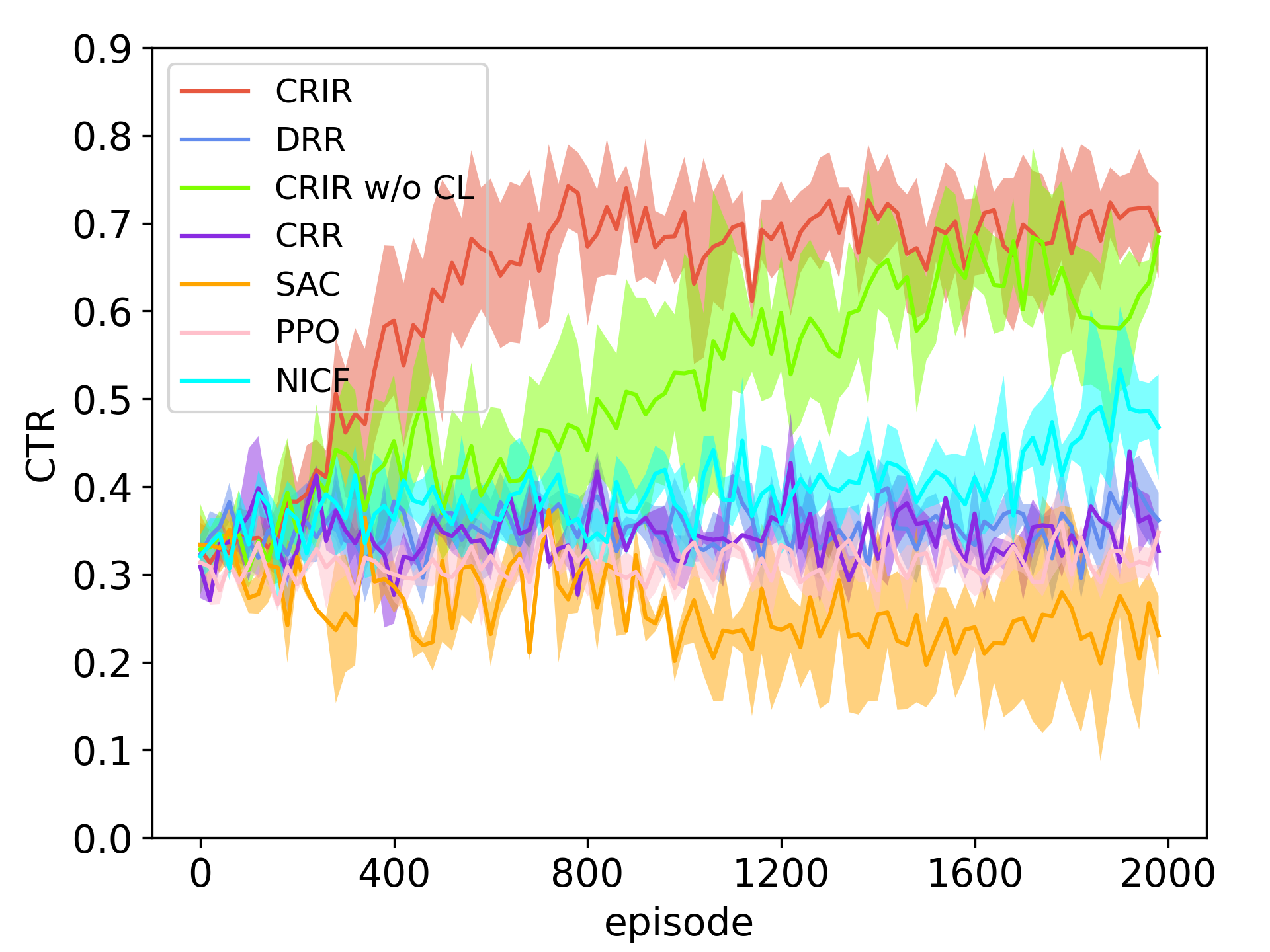}}
	\hfill
	\subcaptionbox{Episode rewards for PRCL frequency study }{\includegraphics[width = 0.33\textwidth]{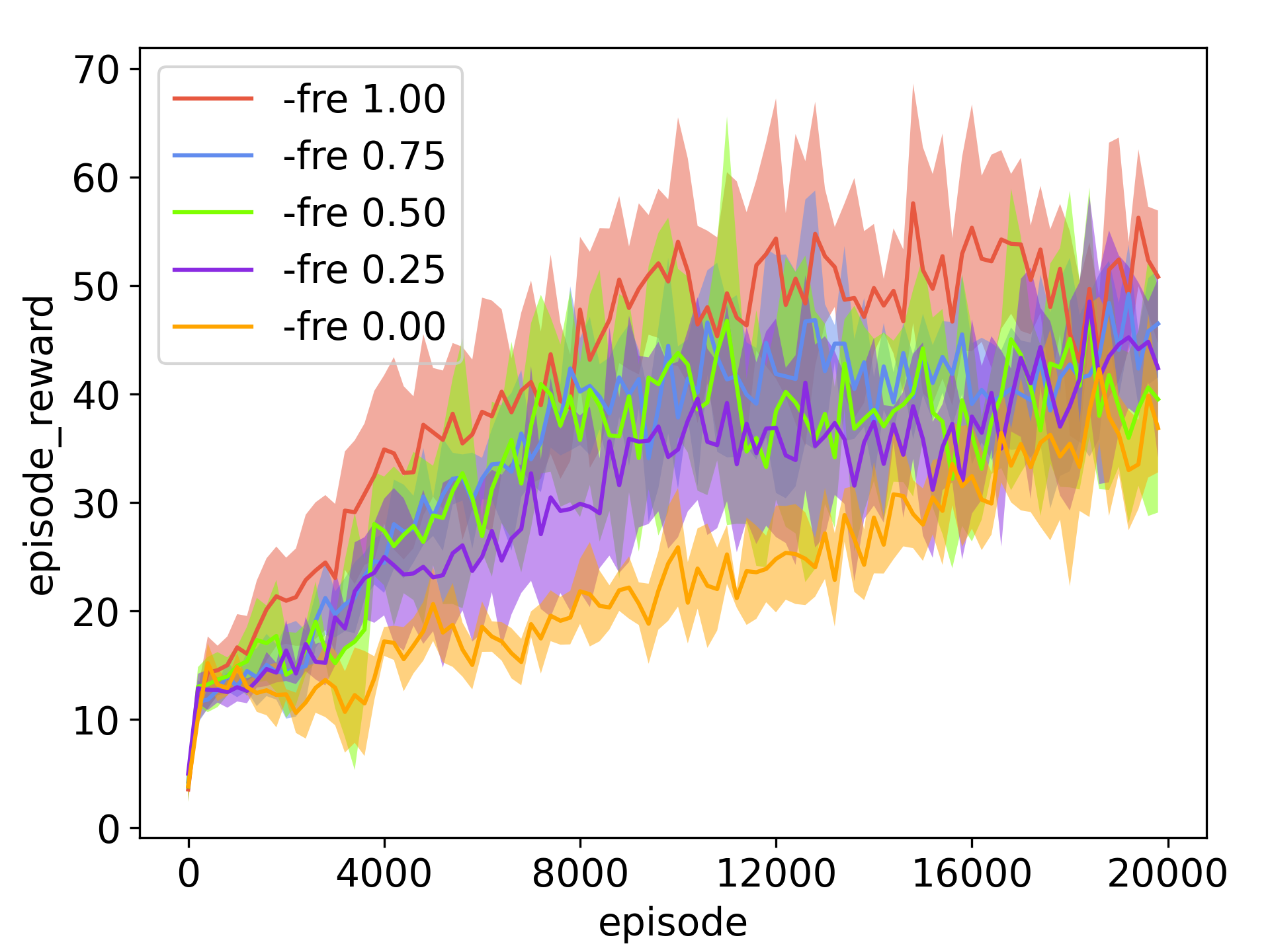}}
	\hfill
	\subcaptionbox{CTR for PRCL frequency study}{\includegraphics[width = 0.33\textwidth]{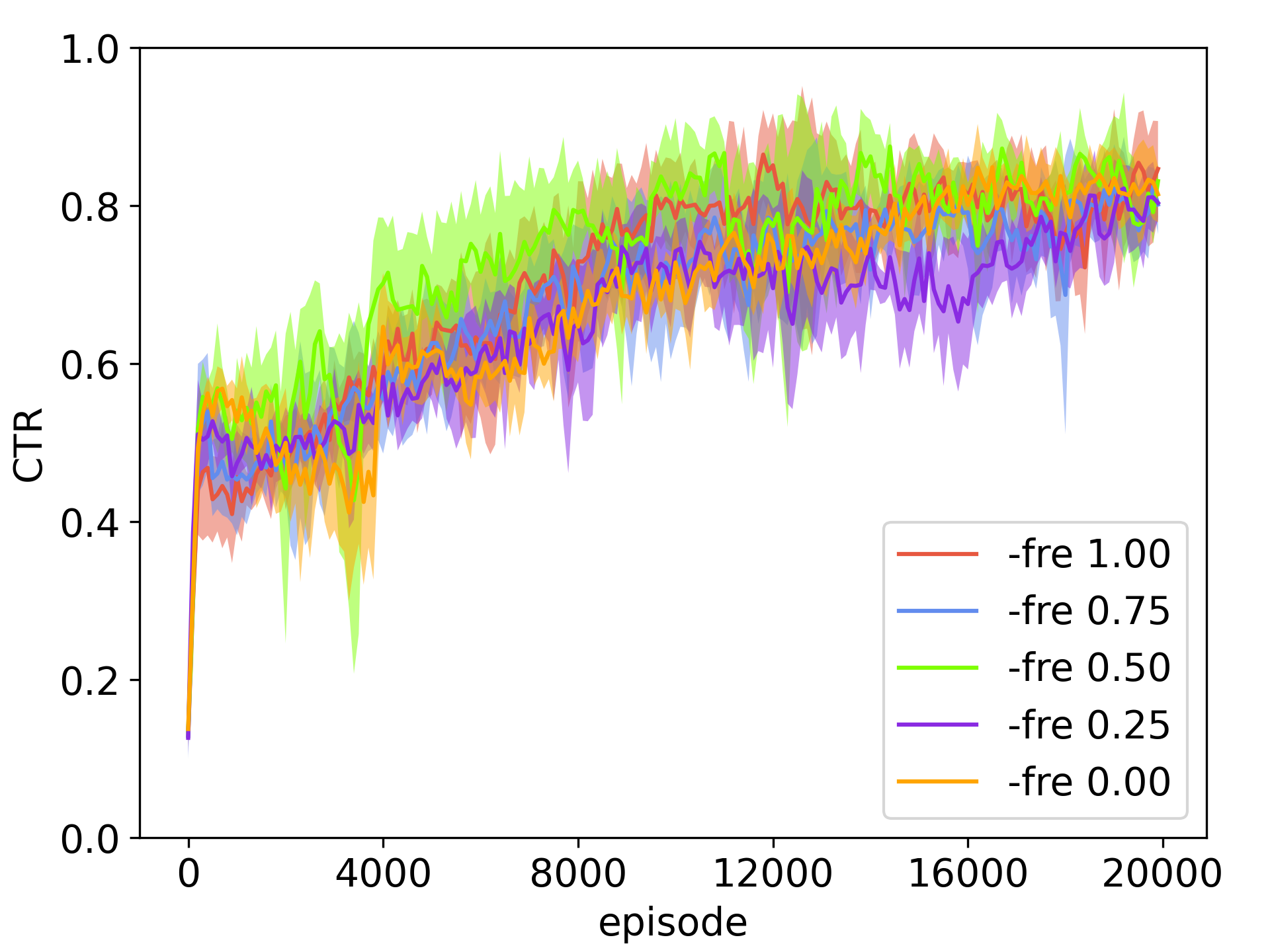}} 
    \caption{Performance for the proposed, ablation and baseline methods in cold-start setting. Each curve in the graph is repeated for 5 times and 95\% confidence intervals are depicted.  (a)-(d) are the results for RQ1, (e) and (f) are for RQ2-1.}\label{RQ12}
\label{fig:label}
\end{figure*}

\subsubsection{Recommendation Environment} 
Traditional recommendation datasets are too sparse to evaluate the interactive recommender systems \cite{Matthew}. Because instant feedback is demanded at every timestamp in interactive settings. Dataset can hardly reflect this. So we use \textbf{Virtual-Taobao} \cite{Virtual-Taobao} and a dataset-oriented \textbf{simulator based on ML-1M} \footnote{\url{https://grouplens.org/datasets/movielens/1m/}} to evaluated CRIR and baseline methods. These simulators will generate a reward signal towards every recommendation reflecting the performance, which satisfy our problem settings. Specifically, we add some dynamic features, like shifting interest, to the ML-1M-based simulator. This is intended for verifying whether experimental methods could perfectly catch dynamic information in recommendation environment.

To fully investigate the sample efficiency of each model, we conduct our experiment \textbf{in totally cold-start settings}, which means all representation parameters are randomly initialized. The model with superior sample efficiency can quickly learn features of users and items from scratch.  

\subsubsection{Evaluation Metrics}
We use two widely used metrics in IR: \textbf{Cumulative Reward} $\sum_t r_t$ in an episode and \textbf{Click Through Rate (CTR)} as our evaluation metric, following previous IR works \cite{CIRS,LSER, Matthew}. Here CTR is denoted as the proportion of positive rewards among all rewards in an episode. Positive reward is denoted as those rewards greater than 0 in both of the two simulation environment.

Sample efficiency is measured through the training effect within the same quantity of data \cite{mai2022sample}. So we use line chart rather than static table to fully display experimental result at every episode. 

There are two reasons why IR cannot be evaluated by list-wise accuracy indicators such as NDCG@K, HR@K. One reason is that precision-based metrics cannot reflect the performance of decision tasks \cite{Matthew}. Another reason is that IR usually applies generative recommendation method rather than scoring-and-ranking method.

\subsubsection{Baselines}
The reason why we chose these baselines is in Section \ref{related work}.
\begin{itemize}
    \item \textbf{SAC}, named Soft Actor Critic, utilized action distribution construct an entropy to constrain action space.
    \item \textbf{CRR}, named Critic Regularized Regression, is a model-free RL method that improve sample efficiency by regularizing weights for policy learning.
    \item \textbf{PPO}, named Proximal Policy Optimization, optimizes a surrogate objective function with gradient ascent while limiting the policy update size to ensure stability.
    \item \textbf{DRR} explored some feasible state representations and investigated a basic generative paradigm applying DDPG for IR. 
    \item \textbf{NICF}, named Neural Interactive Collaborative Filtering, utilize Q-learning and multi-channel transformer to enhance the exploration policy.
    \item \textbf{CRIR w/o CL (Ablation Study)} is our CRIR method without our PRCL approach. It only enhanced the state representation network . This experiment is conducted to verify the contribution of PRCL. 
\end{itemize}

Similar to the DRR, CRIR also use DDPG as implementation backbone and utilize a generative recommendation paradigm. So through the comparison between DRR and CRIR w/o CL, the contribution of the designed state representation network could be verified.


\subsection{Overall Performance and Ablation Study (RQ1)}

\textbf{We first make observations on Virtual-Taobao environment}. Figure~\ref{RQ12} (a) shows the cumulative reward metric. Our CRIR approach outperforms the others in the Virtual-Taobao environment under cold-start settings. It takes the lead in finding a good recommendation policy at around 8000-th episode, while the others have not reached this level within 20000 episodes. The ablation study between whole CRIR and CRIR w/o CL, as well as that between CRIR w/o CL and DRR, demonstrate the contribution of PRCL and the state representation network, separately. But the improvement is slight by only use the representation network. The CRIR w/o CL method performs better than DRR, SAC, CRR, PPO and NICF in early stage, but fails to keep up with CRIR, and gradually declines to the same with the others. Its representation structure helps capture the user’s interest initially but fails to make further progress in subsequent episodes. As CTR metric depicted in Figure~\ref{RQ12} (b) shows, most of the models finally rise up to around 0.8. Note that the reward signals for Virtual-Taobao are greater or equals to 0, which makes high CTR scores easy to achieve. So through the comparison between cumulative reward and CTR, we can know that although most baselines finally reach the same level with CRIR at CTR metric, they get less high-reward actions than CRIR. Baseline methods expect CRIR w/o CL still suffer from sampling inefficiency before 6000-th episode. SAC can scarcely rise but sometimes succeed in CTR metric. PPO fails to learning a correct policy in cold-start settings. The reason for this is that on-policy methods like PPO could hardly filter unimportant or blured transitions from cold-start representations. These on-policy methods usually require well pre-trained representations. NICF is designed specifically for discrete action space initially, it seems not compatible with continuous environment like Virtual-Taobao. 

\textbf{Then we make observations on ML-1M-based environment}. Figure~\ref{RQ12} (c) shows cumulative reward metric, CRIR  outperforms the best on ML-1M oriented simulator. CRIR w/o CL converges slower than CRIR but outperforms all the other methods. Methods except PRCL , CRIR w/o CL and NICF fails within 2000 episodes in this simulator. One reason is that the dynamic features of user interest change quickly in the simulator. The other reason lies in the cold-start setting. These methods do not have enough sample efficiency to find valid policy in such settings. The CTR metric depicted in Figure~\ref{RQ12} (d) seems very consistent with the cumulative reward metric shown in Figure~\ref{RQ12} (c). The reason for this phenomenon is that the simulator returns rewards ranged from -1 to 1, with positive value roughly equal with negative values.

In summary, experiments conducted in two environments demonstrate the effectiveness of CRIR in improving sample efficiency. The comparison between CRIR w/o CL, DRR and CRIR confirms the utility of CRIR’ state representation network and PRCL method.

\subsection{Contribution Quantitative Study (RQ2)}\label{RQ2}

\begin{figure*}[ht]
\centering
        \subcaptionbox{(RQ2-2) Different coefficient strategy}{\includegraphics[width=0.33\linewidth]{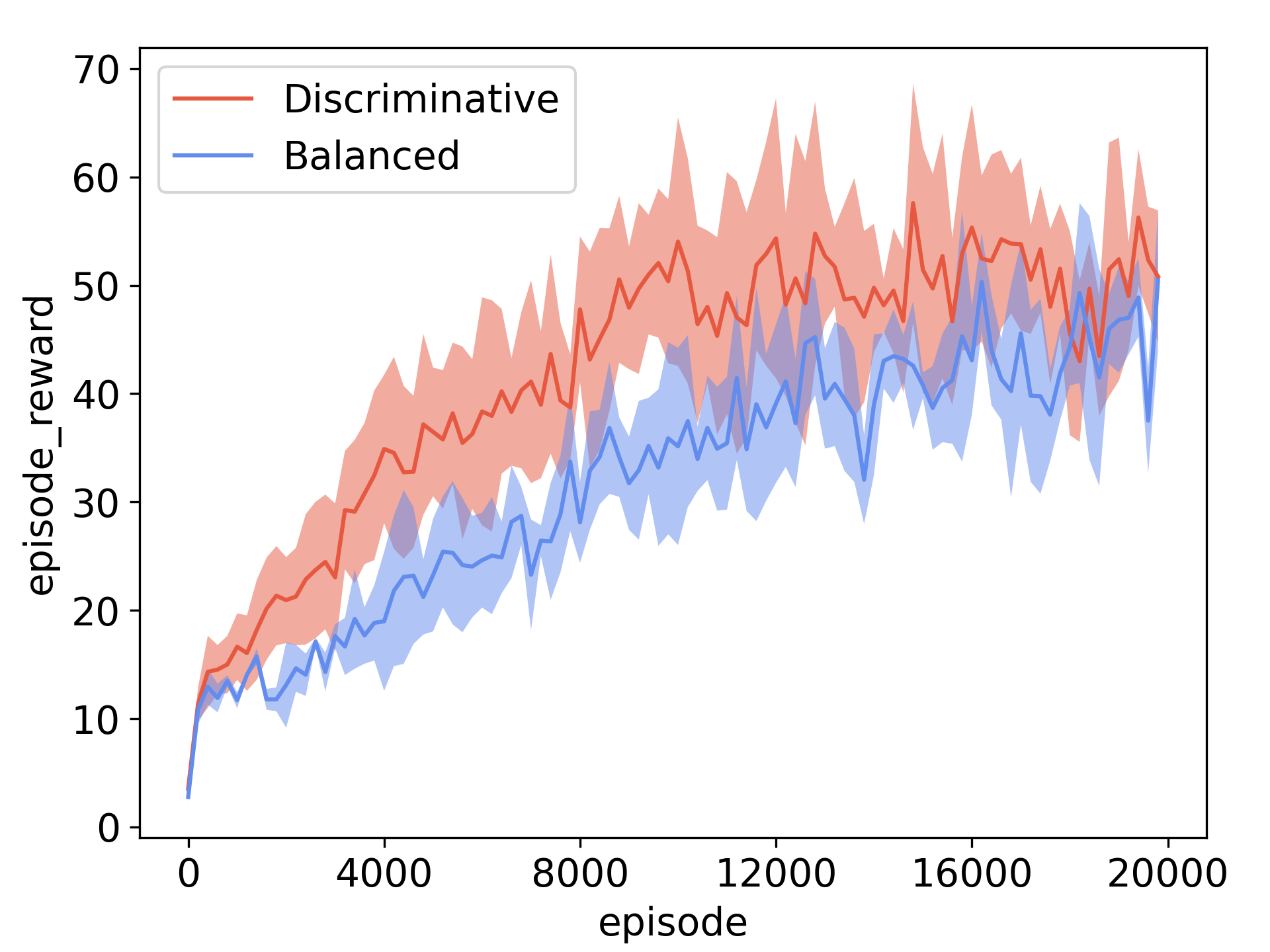}}
        \hfill
        \subcaptionbox{(RQ3-1) Different sampling mechanism}{\includegraphics[width=0.33\linewidth]{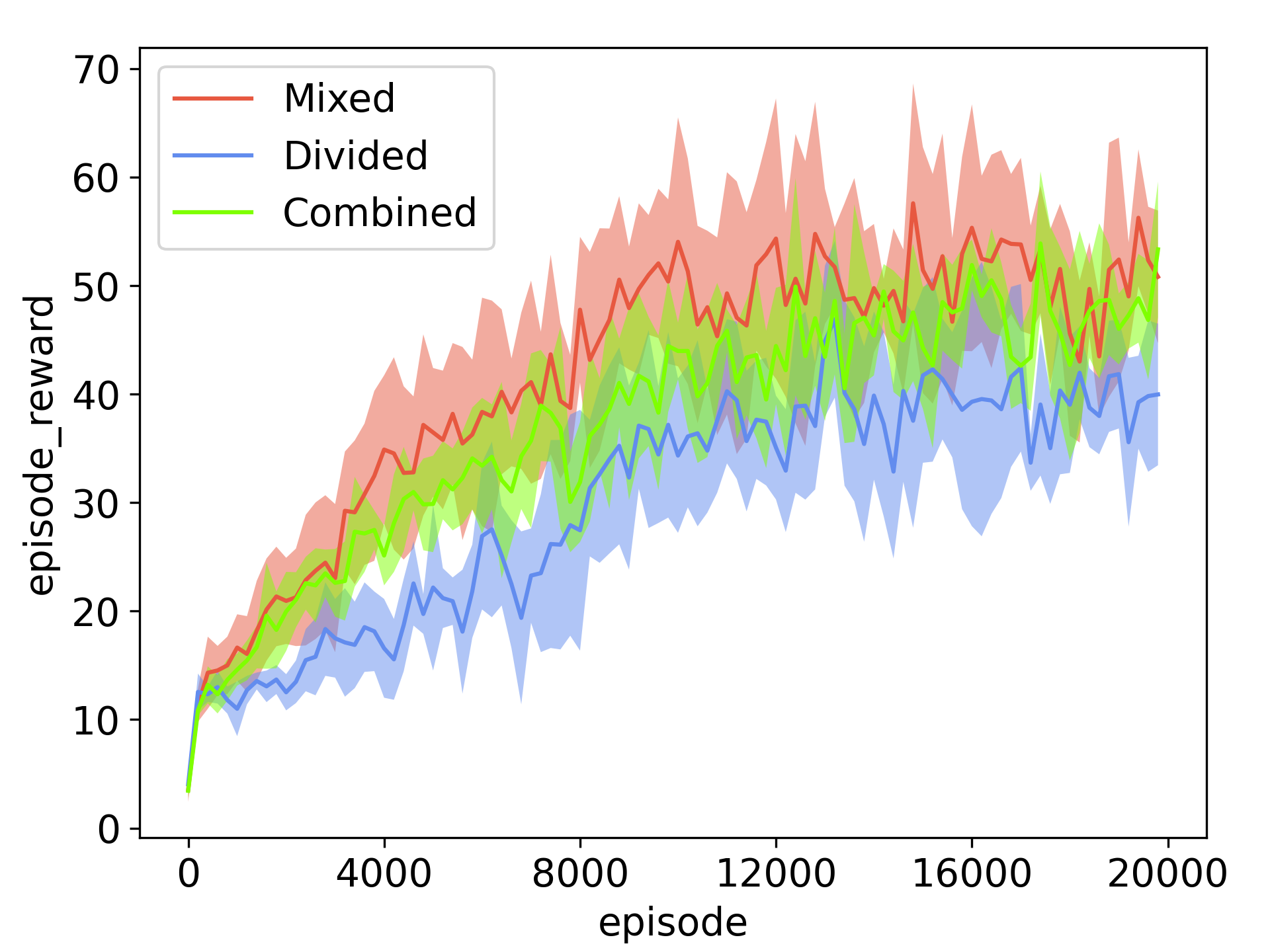}}
        \hfill
        \subcaptionbox{(RQ3-2) Different training mechanism}{\includegraphics[width=0.33\linewidth]{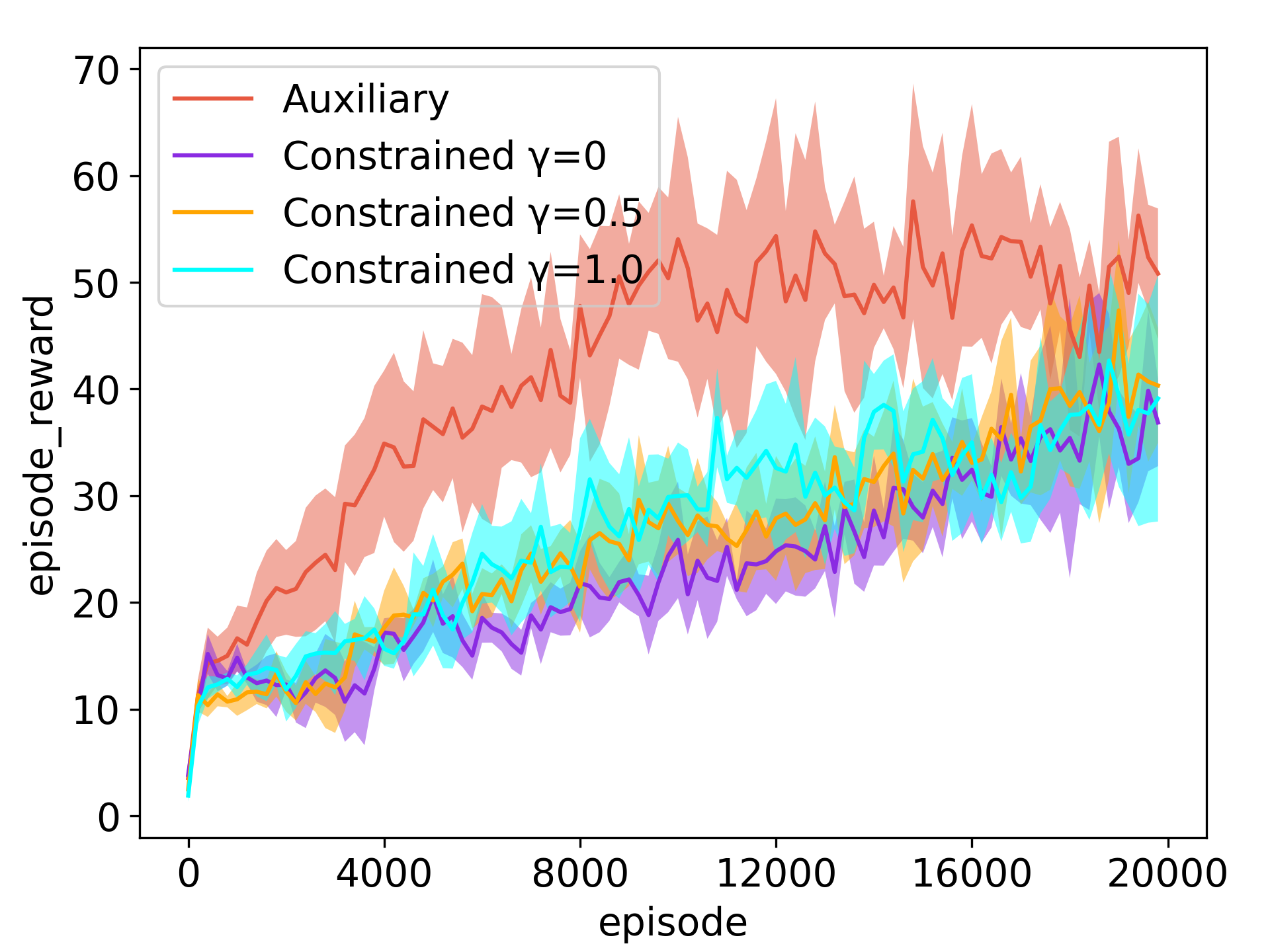}}
    \caption{Study on coefficient strategy, data sampling and agent training mechanism of PRCL. Each curve is repeated for 5 times and 95\% confidence intervals are depicted.}
\label{addexp}
\end{figure*}

We conduct two quantitative contribution studies on two key factors of CRIR to study their detailed contributions. The first experiment studies quantitative research on different frequency of the PRCL. The frequency of PRCL is denoted as the ratio of the times PRCL conducted in that of the RL. The second experiment studies the significance of the discriminative coefficients $1/\sqrt{R_u(h_k)}$ in equation Eq. (\ref{rcl}). We make these two experiments on Virtual-Taobao.

For the first experiment (\textbf{RQ2-1}), we set the PRCL frequencies in $\{0, 0.25, 0.5, 0.75, 1.0\}$. Episode reward and CTR are shown in Figure~\ref{RQ12} (e) and (f) separately. The sample efficiency is boosted with the increase of PRCL frequency. The increment of performance seems not linear. The difference between 0.25 and 0.5 is much larger than that between 0.5 and 0.75. It states that PRCL can effectively improve sample efficiency. But the increment has a limit while increasing the frequency. And PRCL is more capable of optimizing hard metrics like episode reward, than easy metrics like CTR.

For the second experiment (\textbf{RQ2-2}), we use balanced coefficients $w$ to replace the $1/\sqrt{R_u(h_k)}$ in equation Eq. (\ref{rcl}) as baseline method. To guarantee the same average intensity for contrastive learning, we set all the coefficients to $w=(1/\lfloor T/2\rfloor)\sum_{i=2}^{\lfloor T/2\rfloor}(1/\sqrt{i})\approx 0.3183$ where $T=50$ is the max sequence length for state representation. The result can be seen in Figure~\ref{addexp}(a). As the result shown, the discriminative coefficient strategy performs better than the balanced one. The balanced strategy performs approximately the same with the PRCL with learning frequency of 0.25 in Figure~\ref{RQ12}(e). This demonstrates the utility of the proposed discriminative coefficients in PRCL.

\subsection{Sampling and Training Mechanism Study (RQ3)}

We will verify our proposed data sampling and training mechanism mentioned in Section~\ref{dataaug} \textbf{(RQ3-1)}. The proposed data sampling mechanism utilizes two batches of interaction data for PRCL --- one is the batch planed to train DRL immediately (sampled by PER strategy) while the another is randomly sampled from the replay buffer. This sampling strategy is named as \textbf{Mixed Mechanism}. Accordingly, we may consider other two mechanisms --- totally sampling randomly from the buffer, or just using the data planed for DRL training. We name them \textbf{Divided Mechanism} and \textbf{Combined Mechanism} separately.

Considering that our PRCL task is conducted independently with the DRL, we also study the dependent way to conduct PRCL \textbf{(RQ3-2)}. In DRL task, our state representation network is updated by value function (critic network) in DRL. So it means that the Positional Weighted InfoNCE Loss is added to the loss of the value function as a constraint. In this way the loss of value function is formulated as:
\begin{equation}\label{constrain}
\mathcal{L}=\frac{1}{2}{\delta}^{2}+\gamma*\mathcal{L}_{PRCL}
\end{equation}
where $\delta$ is the TD-error in DRL, $\gamma$ is hyper-parameter that controls the strength of PRCL, and $\mathcal{L}_{PRCL}$ is defined in Eq~(\ref{rcl}). We name this training mechanism as \textbf{Constrained Mechanism}. We choose $\gamma \in \{0, 0.5, 1.0\}$. Conversely, we name CRIR's training strategy as \textbf{Auxiliary Mechanism}.

As shown in Figure~\ref{addexp}(b), the Mixed Mechanism performs the best among all data sampling strategies. The comparison between Mixed, Divided and Combined Mechanism demonstrates the effectiveness of our Mixed Mechanism. It shows that representation will be learnt better by utilizing DRL training samples along with some extra samples. As shown in Figure~\ref{addexp}(c), the $\gamma$ value have little effect on the performance. PRCL seems to have no improvement in Constrained Mechanism. This demonstrates the effectiveness of our Auxiliary training strategy.

\section{Conclusion}
This paper states that sample inefficiency is a tricky problem that hinders the development of IR. 
Inspired by contrastive learning in traditional recommendation paradigm, we propose Contrastive Representation for Interactive Recommendation (CRIR), which contains a state representation network and Preference Ranking Contrastive Learning (PRCL). These two methods could help the agent learns better representations. Then the sample efficiency is improved according to the DRL Representation Consensus. 
Different from precious works, we apply an auxiliary contrastive learning task in parallel with the main DRL task. We also adopt an data sampling strategy to ensure the different optimization goals will not be conflicted.
Extensive experiments have verified the effectiveness of the the proposed CRIR.

\section*{Acknowledgments}
Zhiyong Feng is the corresponding author. This work was supported by the National Natural Science Foundation of China (NSFC) (Grant Numbers 62372323, 62422210, 62276187). 

\bibliography{aaai25}

\clearpage
\appendix
\section{Problem Formulation}
\label{preliminaries}
\subsection{Markov Decision Process}

In particular, reinforcement learning-based recommendation learns from interactions. It is generally formulated as a Markov Decision Process (MDP). We use a quintuple \begin{math}
  \mathcal{M}=(\mathcal{S}, \mathcal{A}, \mathcal{T}, \mathcal{R}, \gamma)
\end{math}
to describe the process, where \begin{math}\mathcal{S}\end{math} and \begin{math}\mathcal{A}\end{math} represent for the state space and action space, \begin{math}\mathcal{T}\end{math} represents for the transition probability from observed state to the next states, \begin{math}\mathcal{R}\end{math} is the set of possible reward and \begin{math}\gamma\end{math} is the discounted factor in RL. At timestamp \begin{math}t\end{math}, the policy of RL will initially make an action \begin{math}a_{t} \in \mathcal{A}\end{math} according to the current observed state \begin{math}s \in \mathcal{S}\end{math}. Then the reward \begin{math}r \in \mathcal{R}\end{math} is returned, and the state is transferred to \begin{math}s_{t+1}\end{math} according to transition probability \begin{math}
T(s,a,s^{'})=P(s_{t+1}=s^{'}|s_{t}=s,a_{t}=a)
\end{math}.

We denote the set of all possible users as \begin{math}\mathcal{U} \subset \mathbb{R}^{F_\mathcal{U}}\end{math}, where \begin{math}F_{u}\end{math} is the feature number of the user. The set of all possible item is denoted as \begin{math}\mathcal{I} \subset \mathbb{R}^{F_\mathcal{I}}\end{math}, where \begin{math}F_\mathcal{I}\end{math} is the feature number of the user. The interaction sequence for a user  \begin{math}u\end{math} can be described as \begin{math}\mathscr{L}_{u}=\{ S_{u}^{1}, S_{u}^{2}, \cdots, S_{u}^{|\mathscr{L}_{u}|} \} \end{math}, where each \begin{math}S_{u}^{k} \in \mathcal{S} \end{math} is the \begin{math}k\end{math}-th recorded state. This process can be cast as a reinforcement learning problem, whose key components are summarized as follows:
\begin{itemize}
\item{\textbf{Environment}}: RL agents works in the environment where states and rewards are generated. In this work, the environment must take the responsibility to record the interaction history \begin{math}\mathscr{L}_{u} \end{math} for current active user.
\item{\textbf{State}}: The state in our work contains the information of the current active user \begin{math}U \in \mathcal{U}\end{math}. It can be represented by an embedding vector \begin{math}\mathbf{u}_{t} \in \mathbb{R}^{D}\end{math} with embedding size \begin{math}D\end{math} at timestamp \begin{math}t\end{math}. It also includes user interaction history \begin{math}\mathscr{L}_{u} \end{math}. Each \begin{math}S_{u}^{k} \in \mathcal{S} \end{math} contains interacted items \begin{math}I_{t} \in \mathcal{I}\end{math} and feedback \begin{math}R_{t} \in \mathbb{R}\end{math} from user. Each item and its related feedback can be represented by \begin{math}\mathbf{i}_{t} \in \mathbb{R}^{D}\end{math}. The state at a specific timestamp \begin{math}t\end{math} will be approximated to \begin{math}\mathbf{s}_t\in \mathbb{R}^{D_S}\end{math} through all kinds of transformation.
\item{\textbf{Action}}: The system makes the action \begin{math}a_{t} \in \mathcal{A}\end{math} at time \begin{math}t\end{math} to recommend items to current user. The vector \begin{math}\mathbf{e}_{a_{t}} \in \mathbb{R}^{D}\end{math} is an vector with dimension \begin{math}D\end{math}, sharing the same space with representation of items. 
\item{\textbf{Reward}}: The current active user returns feedback as a reward score \begin{math}r_{t}\end{math} reflecting its satisfaction after receiving a recommended item.
\item{\textbf{State Transition}}: After the agent makes an action \begin{math}a_{t}\end{math} and the user gives a reward \begin{math}r_{t} \in R_t\end{math}, the state \begin{math}s_{t}\end{math} will be updated to \begin{math}s_{t+1}\end{math} according to a state transition probability \begin{math}T(s_{t+1}=s^{'}|s_{t},a_{t})\end{math}.
\end{itemize}

If the environment is discrete (means that user and items are in ID forms), the received action \begin{math}a_{t}\end{math} will degrade from \begin{math}\mathbb{R}^{F_\mathcal{I}}\end{math} into \begin{math}id \in \mathbb{N} \end{math}, where $id$ is the ID of a possible item. The action \begin{math}a_{t}\end{math} will be made by cosine similarity of embedding with that of \begin{math}\mathbf{i}_{t}\end{math}, which is \begin{math}
    a_{t} = \mathop{\mathrm{argmin}}\limits_{id \in \mathcal{I}} \frac{\mathbf{e}_{a_t}^{\rm T} \cdot E(id)}{\|\mathbf{e}_{a_t}\|\cdot \|E(id)\| } 
\end{math}, where \begin{math}E:\mathbb{N}\rightarrow\mathbb{R}^D\end{math} is the embedding module transferring $id$ to embedding.

\label{3.1}If the environment is continuous(means that users and items are in feature vector form, like Virtual-Taobao \cite{Virtual-Taobao}), the user feature \begin{math}U \in \mathbb{R}^{F_\mathcal{U}}\end{math} and item feature \begin{math}I \in \mathbb{R}^{F_\mathcal{I}}\end{math} are also encoded into \begin{math}\mathbf{u}_{t}, \mathbf{i}_{t} \in \mathbb{R}^{D}\end{math} with embedding size \begin{math}D\end{math} at timestamp \begin{math}t\end{math} separately. The received action is formed as \begin{math}\mathbb{R}^{F_\mathcal{I}}\end{math}. The $E:\mathbb{R}^{F_\mathcal{I}}\rightarrow\mathbb{R}^D$ serves as the encoding module, transferring sparse feature vectors to dense feature vectors.

\subsection{DRL and State Representation}\label{3.2}

Current DRL methods usually contain both policy or value function, which are implemented through actor and critic network\cite{RLSurvey}. The policy function is employed to generate actions \begin{math}a_t \sim \pi (s_t)\end{math} , where \begin{math}\pi(\cdot)\end{math} is the policy in RL after state \begin{math}s_t\end{math} in MDP \begin{math}\mathcal{M}\end{math}. While the value function outputs the expected discounted reward \begin{math}
    V^{\pi}(s)=\mathbb{E}_{a_t \sim \pi (s_t)}[\sum_{t=0}^{\infty} \gamma^{t}r(s_t, a_t)|s_0=s]
\end{math}. Therefore, the optimization goal of critic network is defined by mean-square loss $\frac{1}{2}{\delta}^{2}$ of TD-Error $\delta$, where $\delta$ is defined in Eq. (\ref{value function}):
\begin{equation}\label{value function}
    \delta = r(a_t,s_t) + \gamma V^{\pi}(s_{t+1})-V^{\pi}(s_t)
\end{equation}
After optimizing value function, the policy function is optimized by policy gradient, which can be formulated as Eq. (\ref{policy function}).
\begin{equation}\label{policy function}
    \nabla_{\theta} J_{t}(\theta) = \mathbb{E}_{a_t \sim \pi_{\theta }(s_t)}[\nabla_{\theta}\log\pi_{\theta}(a_t|s_t)Q(s_t, a_t))]
\end{equation}
where $Q(s_t, a_t)=\sum_{t=0}^{\infty} \gamma^{t}r(s_t, a_t)$, is the discounted return for a specific action.

When facing complex environment like IR, the information for the current observation need various variables and parameters to describe. Therefore, the state information should be delicately extracted into high level representations to ease DRL training. In this case, a state representation network is usually demanded to encode all the state information affecting the agents to make decisions into a representation \begin{math}\mathbf{s}_t \in \mathbb{R}^{D_S}\end{math} at timestamp \begin{math}t\end{math}, where \begin{math}D_S\end{math} is the dimension of the representation. In this case, the state representation network is regarded as the shared prefix layers for both actor and critic network. And its parameters are updated through the gradient passed from DRL components. Usually we update state representation network together with value function $V^{\pi}(s)$ to stabilize training.

Experience Replay mechanism is widely utilized in off-policy RL methods \cite{RLSurvey}. It utilizes a replay buffer to storage agent's past interaction experience to help agent review past knowledge. It greatly helps the agents improve its performance. Commonly, each experience consists of a tuple $(s, a, r, s', DONE)$, where: $s$ is the current state, $a$ is the action that was taken preciously, $r$ is the reward signal received from environment, $DONE$ is an boolean indicating whether the interaction is terminated at this state. While training, the agent randomly samples a batch of experience from the replay buffer to update policy and value function. Different experience sample strategies will greatly affect the performance of model's training (The concept of sample efficiency comes from here, which means the average contribution that one experience can make). There has been a variety of experience replay methods \cite{AER,SER,HER,DER,LSER}. In this paper we use Priority Experience Replay \cite{PER}.

The overall CRIR training pseudo code implemented by DDPG can be described by Algorithm \ref{alg:ddpg}.

\begin{algorithm}
\caption{CRIR training procedures in DDPG backbone}\label{Algorithm 1}
\label{alg:ddpg}
\begin{algorithmic}[1]
\REQUIRE initial on-policy, off-policy DRL parameters $\theta, \theta'$, momentum update factor $\tau$,  state representation parameters $\phi$, embedding or encoding parameter $\omega$, empty replay buffer $\mathcal{D}$
\ENSURE  final policy $\pi_{\theta}$
\STATE Initialize all learnable parameters
\FOR{$episode=1, \dots, M$}
    \STATE Receive initial observation state $s_1$, $step\gets 0$
    \WHILE{not $done$}
        \STATE Obtain state representation $e_{s_t}$ from $s_t$.
        \STATE Select action $a_t \sim \pi_{\theta}$ and execute it.
        \STATE Observe reward $r_t$, new state $s_{t+1}$ and $done$.
        \STATE Store transition $(s_t, a_t, r_t, s_{t+1}, done)$ into $\mathcal{D}$.
        \STATE Sample a batch $(s_i, a_i, r_i, s_{i+1}, done)$ from $\mathcal{D}$.
        \STATE Preference Ranking Contrastive Learning to update $\omega$.
        \STATE Update $\theta$ and $\phi, \omega$ through back propagation from RL losses.
        \STATE Momentum update $\theta'\leftarrow \tau\theta+(1-\tau)\theta'$.
        \STATE $s_{step+1}\gets s_{step}$, $step\gets step+1$.
    \ENDWHILE
\ENDFOR
\end{algorithmic}
\end{algorithm}

\section{PRCL Method Formulation}\label{apdx-2}
We will specifically formulate our PRCL method in this section. The total optimization goal of interactive recommendation task can be formulated as:
\begin{equation}
      {\omega}^*,{\theta}^*=\mathop{\mathrm{argmax}}\limits_{\omega,\theta} \ J_{\omega,\theta}, 
\end{equation}
where $\omega,\theta$ are the parameter sets for state representation and DRL components, separately. $J_{\omega,\theta}$ is the optimization goal, formulated as:
\begin{equation}
     J_{\omega,\theta}={\sum_{u=1}^{|\mathcal{U}|}\sum_{t=1}^{T_u}\mathbb{E}_{a_{t;\theta}}[r(s_{u,t;\omega},a_{u,t;\theta})]},
\end{equation}
where $\mathcal{U}$ is the user set, $T_u$ is the interaction length for user $u$, $s_{t;\omega}\in \mathbb{R}^{D_S}$ is the representation for user state $s$ at timestamp $t$ with parameter set $\omega$.  $r(s_t,a_t)$ is the reward returned from environment while taking action $a_t$ at state $s_t$. The action $a_{t;\theta}\in \mathbb{R}^D$ is actually the representation of the chosen item to be recommended. The mathematical expectation could be expanded as:
\begin{equation}\label{extended expectation}
J_{\omega,\theta}=\sum_{u=1}^{|\mathcal{U}|}\sum_{t=1}^{T_u}\sum_{k}^{|\mathcal{A}_{s_{u,t}}|}P(s_{u,t;\omega},a_{u,t,k;\theta})r(s_{u,t;\omega},a_{u,t,k;\theta}),
\end{equation}
where $\mathcal{A}_s$ is the possible action set for state $s$, $P(s_{u,t},a_{u,t,k})$ is the joint probability for the agent taking action $a_{u,t,k}$ at state $s_{u,t}$. 

Since Eq~(\ref{extended expectation}) is hard to optimize, we can formulate a lower-bound for it by applying an inequality:
\begin{align}
&J_{\omega,\theta}\geq\sum_{u=1}^{|\mathcal{U}|}\sum_{t=1}^{T_u}\sum_{k}^{|\mathcal{A}_{s_{u,t}}|}\ln{P(s_{u,t;\omega},a_{u,t,k;\theta})}+ \\
&\sum_{u=1}^{|\mathcal{U}|}\sum_{t=1}^{T_u}\sum_{k}^{|\mathcal{A}_{s_{u,t}}|}[\ln{r(s_{u,t;\omega},a_{u,t,k;\theta})}+1].\label{DRLTask}
\end{align}

The term Eq~(\ref{DRLTask}) contains short-term reward and interaction length information , which could be optimized by main DRL tasks through Eq~(\ref{policy function}). So we do not consider Eq~(\ref{DRLTask}) in our PRCL. Since our method does not involve user-side disposals, so we only formulate it for one single user. Finally, we get a new maximization goal for PRCL task:
\begin{equation}
J_{u;\omega}=\sum_{t=1}^{T_u}\sum_{k}^{|\mathcal{A}_{s_t}|}\ln{P(s_{t;\omega},a_{t,k})}.
\end{equation}
We can transfer it into conditional probability:
\begin{equation}\label{conditional}
J_{u;\omega}=\sum_{t=1}^{T_u}\sum_{k}^{|\mathcal{A}_{s_t}|}\ln{P(s_{t;\omega})}+\ln{P(a_{t,k}|s_{t;\omega})}.
\end{equation}
As mentioned in section~\ref{4.2}, we model the representation $s_{t;\omega}$ as Eq~(\ref{SR}). We rewrite it here:
\begin{equation}\label{rewrite state}
    s_{t;\omega}= (\sum_{\tau=1}^{t}w_{\tau} h_{\tau}) \oplus (\frac{1}{t}\sum_{\tau=1}^{t}u_{t}^{\rm T}\otimes h_{\tau}).
\end{equation}
where  $w_{\tau}$ is the interest weight generated by activation unit, $h_{\tau}\in \mathbb{R}^{D_B}$ is the behavioral representation. $a_{t,k}$ shares the same space with $h_i$.

We consider the conditional probability $P(a_{t,k}|s_{t;\omega})$ in Eq~(\ref{conditional}). Since it is reasonable for agent to make action according to the current state, we assume that this distribution on actions can be written as Gaussian distribution around state:
\begin{equation}\label{Gaussian}
        P(a_{t,k}|s_{t;\omega})=\frac{\exp(-(a_{t,k}^{\rm T}W_a-s_{t}^{\rm T}W_s)^2)}{\sum_{j=1}^{|\mathcal{A}_{s_t}|}\exp(-(a_{t,j}^{\rm T}W_a-s_{t}^{\rm T}W_s)^2)},
\end{equation}
where $W_a \in \mathbb{R}^{D\times D_c}$ and $W_a \in \mathbb{R}^{D_S\times D_c}$ separately transform the action and state representation to a same projection space $\mathbb{R}^{D_c}$ for contrastive learning. Maximizing Eq~(\ref{conditional}) by bringing in Eq~(\ref{Gaussian}) is equivalent to minimize the following loss function as Eq~(\ref{projection}) shows:
\begin{equation}\label{projection}
        \mathcal{L}_u=-\sum_{t=1}^{T_u}\sum_{k}^{|\mathcal{A}_{s_t}|}\log \frac{\exp(a_{t,k}^{\rm T}W s_{t})}{\sum_{j=1}^{|\mathcal{A}_{s_t}|}\exp(a_{t,j}^{\rm T}W s_{t})},
\end{equation}
where $W\in \mathbb{R}^{D\times D_S}$. However, Eq(\ref{projection}) is also hard to optimize due to the high computational complexity. Moreover, this loss function have not yet consider the  term $\sum_{t=1}^{T_u}\sum_{k}^{|\mathcal{A}_{s_t}|}\ln{P(s_{t;\omega})}$ in Eq~(\ref{conditional}). So we make simplification and optimization for it.

Our simplification idea is to choose one representative action to alternate computation on all possible actions in $\mathcal{A}_{s_t}$. As we can see in Eq~(\ref{rewrite state}), the behavior $h_i$ with larger interest weight value $w_i$ can mostly represent the current state and reflect user's interest points. So we utilize a representative item to approximate the state $s_t$. Based on the idea of utilizing item to represent state, we use $h^{*} =h_i,\ i=\mathop{\mathrm{argmax}}\limits_{i\leq t}\ w_i$ to approximate $s_t$.
Similarly, we also use behavior representation to alternate the action $a_{t,k}$. But it has some differences with alternating the state. We desire to decrease the computational complexity. The Eq~(\ref{projection}) will enlarge the representation distances between all possible actions, which is obviously unnecessary. We just need to enlarge those distances between high reward actions and low reward actions and the $\mathcal{L}_u $ could be optimized too. So we consider using a randomly selected behavior which user is interested in to replace the entire action calculation loop. Moreover, we use uninterested behaviors set $\mathcal{N}_{s_t}$  which have appeared in user‘s interaction history, to alternate the whole possible action set $\mathcal{A}_{s_t}$ . Through this way the loss is reformed as:
\begin{equation}
         \mathcal{L}_u(t)=-\log \frac{\exp(h^{\rm T}_{k}\cdot h^{*})}{\sum_{n=1}^{\mathcal{N}_{s_t}}\exp(h_n^{\rm T} \cdot h^{*})},
\end{equation}
where $t \leq T_u$ is the current timestamp, $k$ is the chosen behavior index from positive set, $\mathcal{N}_{s_t}$ is the negative behavior set at current state $s_t$. The positive and negative set is generated by ranking through all behaviors by order of $w_i$. The split point is set at the interval of the ranking position.

Our optimization idea is to consider the term $\sum_{t=1}^{T_u}\sum_{k}^{|\mathcal{A}_{s_t}|}\ln{P(s_{t;\omega})}$ in Eq~(\ref{conditional}). Here we also use the strategy which use behavior representation to approximate state representation. This term depicts the state distribution in logarithmic representation space. So it can reflect different importance of different states. So we use $\mathbf{1/\sqrt{R_u(h_k)}}$ to smooth the discrimination, where the $R_u(h_k)$ is the ranking position for $h_t$ among user $u$'s history behaviors, ranked by interest weight $w_t$ in Eq~(\ref{rewrite state}). So the final loss of PRCL for one user is formulated as:
\begin{equation}\label{apdx-rcl}
\begin{aligned}
        \mathcal{L}_u(t)=-\frac{1}{\sqrt{R_u(h_k)}}\log\frac{\exp(h^{\rm T}_k\cdot h^{*})}{\sum_{n}^{{\mathcal{N}}_{s_t}}\exp(h_n^{\rm T} \cdot h^{*})}.
\end{aligned}
\end{equation}

Since this loss involves users' dynamic ranking interests, we name it \textbf{Positional weighted InfoNCE Loss}, and we name the whole contrastive learning process as \textbf{Preference Ranking Contrastive Learning}. And the time complexity of this PRCL method is $O(Ud(U+d))$ where $d$ is the size of representation dimension, $U$ is the user amount and $L$ is the average interaction session length. 

\section{Simulation Environments}\label{Environments}
Traditional recommendation datasets are too sparse or lack necessary information to evaluate the interactive recommender systems \cite{Matthew}. Because instant feedback is demanded at every timestamp in interactive settings. Moreover, one core specialty for interactive recommendation is that it can optimize users' long term satisfaction. Dataset can hardly reflect this. So we use Virtual-Taobao \cite{Virtual-Taobao} and a dataset-oriented simulator to evaluated the proposed and baseline models. These simulators will generate a reward signal towards every recommendation reflecting the performance, which satisfy our problem settings.
\begin{itemize}
    \item{\textbf{Virtual-Taobao}} is a real-time virtual user simulation platform, where the agent recommends items according to users’ dynamic interests. It use pre-trained generative adversarial imitation learning (GAIL) to generate different users with both static and dynamic interests. It is the continuous environment mentioned in the section of 'A. Problem formulation'. We apply it as our prime environment. 
    \item{\textbf{ML-1M Oriented Simulator}.} We design a simulation environment based on dataset ML-1M. By imitating the dynamic features in Virtual-Taobao, we add some interest shifting mechanisms into the environment to verify the adaptation of experimental methods toward dynamic features. The pseudo code of reward strategy of this simulator can be shown in Algorithm \ref{simulator}.. We set a $top\_k$ mechanism for ml-1m, either. This means the system should search for the $top\_k$ nearest items in embedding space as the recommendation result. The maximum interaction length of this simulator is set to 50.
\end{itemize}

The max sequence length for state representation is set to 50. The number of episodes is set to 20000 for Virtual-Taobao and 2000 for ML-1M. All embedding dimensions are set to 100. We use two three-layer neural networks with 128 hidden units for actor and critic separately, the learning rate of which are set to 0.001. All the algorithms in the experiment with actor or critic structure are set in that way. All the methods use SRM as their state representations except for CRIR w/o CL and the proposed method. All discounted factors in RL are set to 0.9. The momentum update parameter is set to 0.001 for all algorithm with target networks. All the models are implemented through PyTorch \cite{pytorch}.

\begin{algorithm}
\caption{Reward Strategy for Dataset-Oriented Simulator (ML-1M)}\label{simulator}
\begin{algorithmic}[1]
\REQUIRE recommended size $top\_k$, recommendation items list $Actions$, dataset rating value mapping $rate:\mathcal{A}\rightarrow \mathbb{R}$, interaction history $\mathscr{L}$
\ENSURE $Reward \in [-1, 1]$
\STATE $Reward\gets -1$
\FOR{$action$ in $Actions$}
    \STATE $reward\gets \mathrm{get\_reward}(action)$
    \IF{$reward>Reward$}
        \STATE $Reward\gets reward$
    \ENDIF
\ENDFOR
\RETURN $Reward$
\STATE 
\STATE \textbf{Function} $\mathrm{get\_reward}$($a$)
\STATE \textbf{\quad if} $\mathrm{rate}(a)==1$ \textbf{or} $a$ \textbf{not in} ML-1M Dataset \textbf{then}
\STATE \textbf{\quad\quad return} $-1$ 
\STATE \textbf{\quad }$score\gets (\mathrm{rate}(a)-1)^2$
\STATE \rm{// penalize repeated action}
\STATE \textbf{\quad if} $a \in \mathscr{L}$ \textbf{ then}
\STATE \textbf{\quad\quad}$score\gets  \mathrm{max}(-1,\mathrm{min}(0.3, 1.1-0.2*\mathscr{L}.\mathrm{count}(a)))$
\STATE \rm{// normalize to} $[-1, 1]$
\STATE \textbf{\quad if} $score>0$ \textbf{then}
\STATE \textbf{\quad\quad}$score\gets score/16$
\STATE \textbf{\quad if} $score<-1$ \textbf{then}
\STATE \textbf{\quad\quad}$score\gets -1$
\STATE \textbf{\quad return} $-1$
\STATE \textbf{End Function}
\end{algorithmic}
\end{algorithm}

\section{A Brief Experimental Study on DRL Representation Consensus: from Gradient Perspective}

\begin{figure*}[ht]
\centering
    \subcaptionbox{Episode reward for Virtual-Taobao}{\includegraphics[width = 0.33\textwidth]{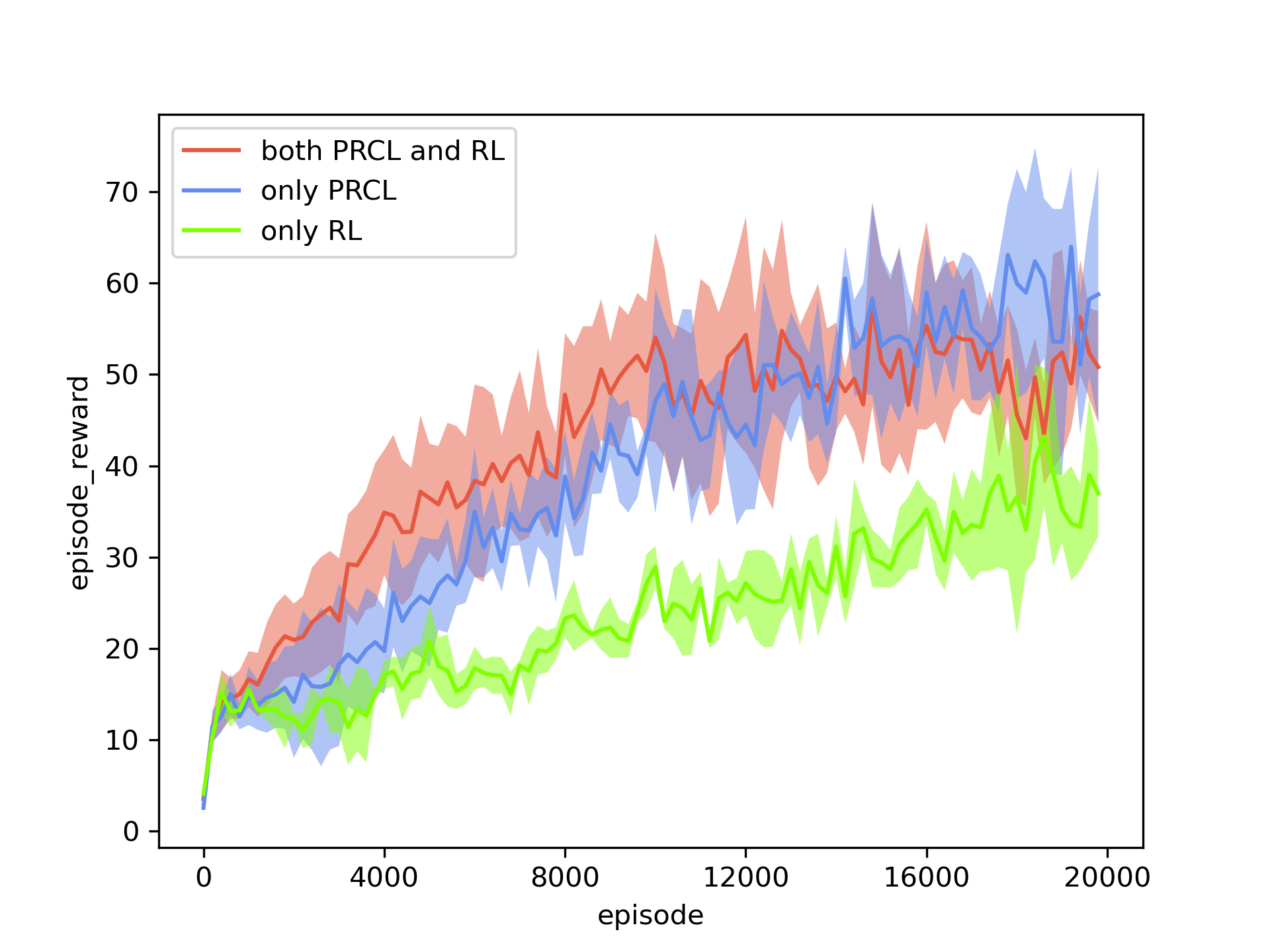}}
    \hfill
    \subcaptionbox{Gradients passed from PRCL.}{\includegraphics[width = 0.33\textwidth]{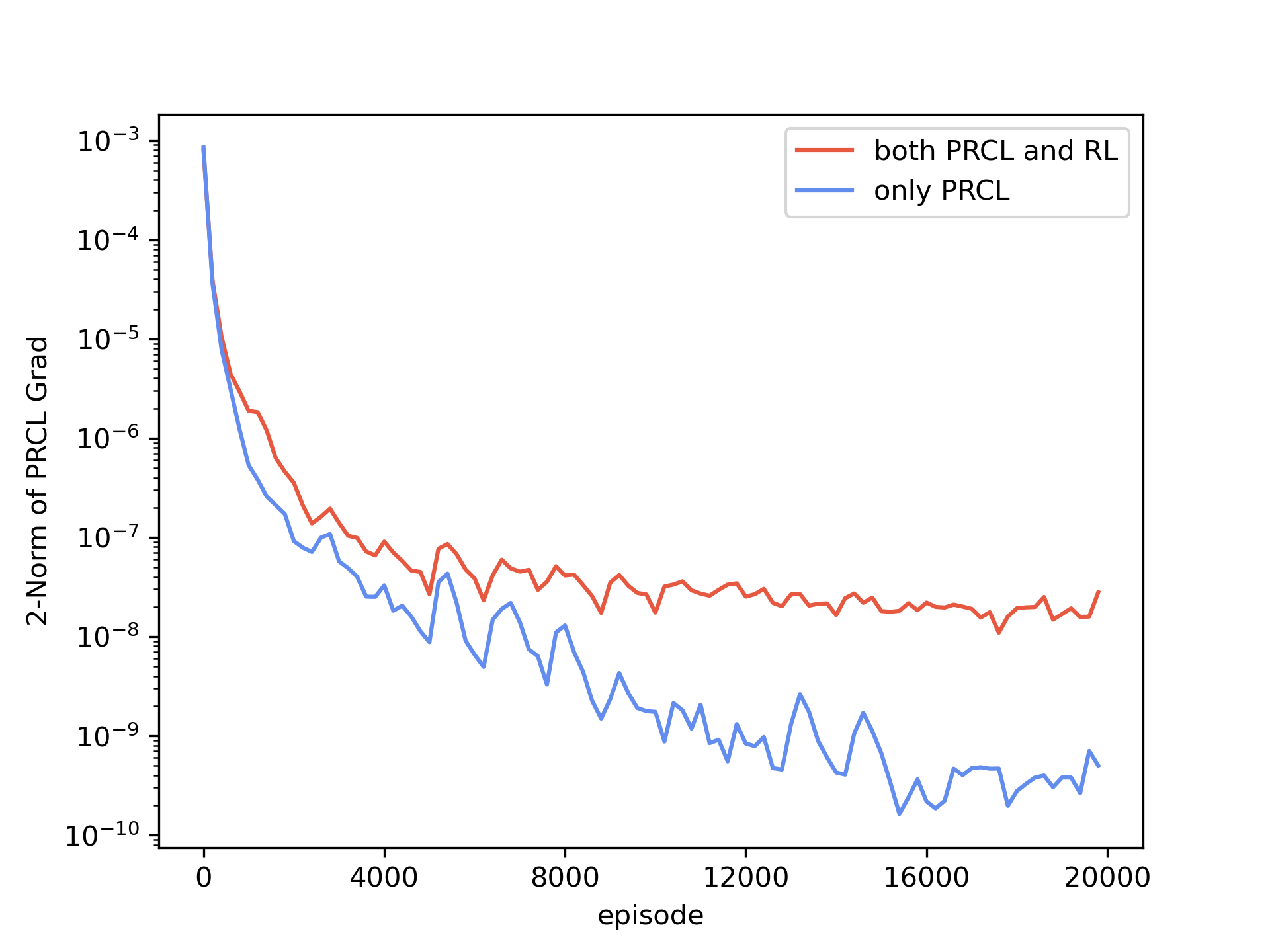}}
    \hfill
    \subcaptionbox{Gradients passed from RL.}{\includegraphics[width = 0.33\textwidth]{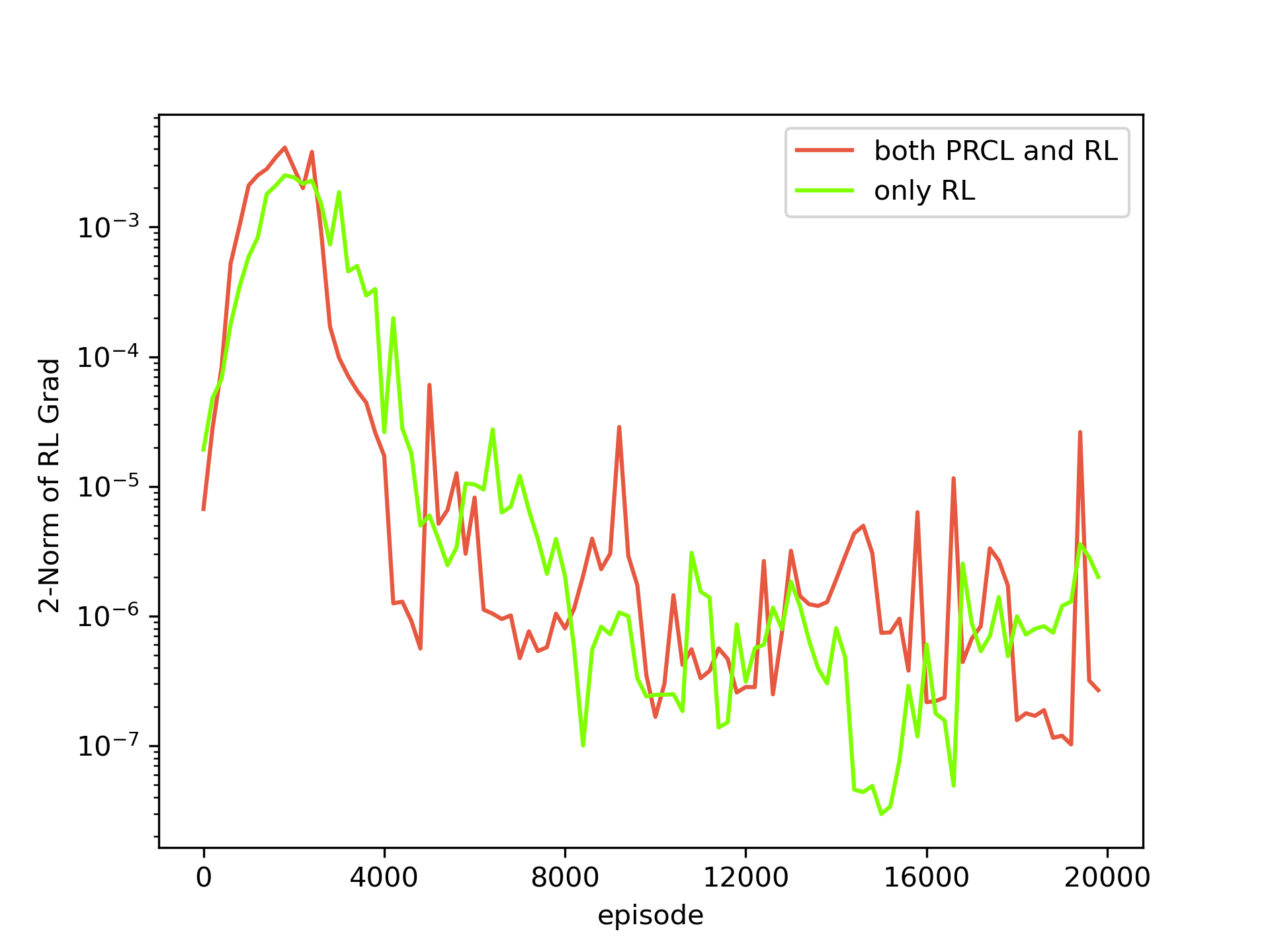}} 
    \caption{Strategies for different gradient updating policy in PRCL and gradient study. Logarithmic axis is used in (b) and (c) to benefit visualization.}
\label{grad}
\end{figure*}

To study how does DRL Representation Consensus work is of vital importance to ensure the basis of our proposed method and certificate its robustness. Our motivation is to observe the results when training representation modules in different ways. If this assumption works, better representation training method will lead to better sample efficiency.

In CRIR, the embedding or encodding layer is updated through two kinds of gradients. One is the gradient backpropagated from losses of DRL, the other is the gradient backpropagated through the PRCL. So there naturally exists three kinds of behavior encoding layer update methods that can be investigated. The first one is to only use gradient from RL algorithm shown in the legend 'only RL' to update the encoding layer. Actually it refers to the CRIR w/o CL method in (RQ1). The second one is to only use gradient from the proposed PRCL method shown in the legend 'only PRCL' to update the encoding layer. The last one is to use both of them shown in the legend 'both RL and PRCL'. It refers to our proposed whole method.

The experimental results are shown in Figure~\ref{grad} (a). As we can see in the result, though just using PRCL to update the parameters, the method 'only PRCL' performs solely fall behind a little with the original one. It exceeds the 'both RL and PRCL' in the end. And the CRIR w/o CL method doesn't catch up with the others. This demonstrates that the PRCL makes good contributions to improving sample efficiency, which is much greater than that of RL components do. And it seems that there exist conflicts between the two approaches to update the embedding. 

So to further study the connections between representation and sample efficiency, we conduct comparison experiments. We record the gradients of a linear layer in the item encoder network during PRCL period in Figure~\ref{grad} (b) and during the parameter updating stage in Figure~\ref{grad} (c). The gradient of the linear layer is a vector $\mathbf{i}_t \in \mathbb{R}^{100}$. So we use $\|\mathbf{i}_t\|_2 $ to reflect the scale of the gradient. Figure~\ref{grad} (b) and Figure~\ref{grad} (c) makes comparison on the gradient generated by the PRCL and the back-propagation of RL separately. The compared curves in Figure~\ref{grad} (c) have similar changing features. So we can know that PRCL do not interfere the gradients at encoding layer back-propagated from RL. But from Figure~\ref{grad} (b) we can acknowledge that the gradients back propagated from RL components may more or less disturb the the learning process of PRCL. Because the gradients from both RL and PRCL doesn't decrease to the level as the gradient only passed from PRCL do. The representation does not learns better with both gradients. It's an amazing but real discovery, which means that the state representation network in DRL doesn't need to trained or fine-tuned by RL losses. Better representation of IR will lead to better sample efficiency.
\end{document}